\documentclass{ws-ijmpa}

\usepackage[super]{cite}
\usepackage{xcolor}
\usepackage[verbose,hypertexnames=false]{hyperref}
\hypersetup{colorlinks=false,allbordercolors=blue,pdfborderstyle={/S/U/W 1}}

\usepackage{pgf}

\usepackage{dcolumn}
\usepackage{bm}

\usepackage{floatrow}
\usepackage{enumitem}

\usepackage{xspace}
\usepackage{multirow}

\usepackage{overpic}
\usepackage{subfig}

\DeclareSubrefFormat{myparens}{#1~(#2)}
\DeclareCaptionListOfFormat{myparens}{#1(#2)}
\graphicspath{{imgs/}}

\newcommand\MeV{\ensuremath{\mathrm{MeV}}}
\newcommand\GeV{\ensuremath{\mathrm{GeV}}}

\begin{document}

\markboth{Yanping Huang}{Discovery of a Glueball-like particle $X(2370)$ at BESIII}

\catchline{}{}{}{}{}

\title{Discovery of a Glueball-like particle $X(2370)$ at BESIII}

\author{Yanping Huang\textsuperscript{\textit a}}
\author{Shan Jin\textsuperscript{\textit b}}
\author{Peng Zhang\textsuperscript{\textit a}}

\address{\textsuperscript{\textit a}Institution of high energy physics, Beijing 100049, People’s Republic of China \\
huangyp@ihep.ac.cn\\
zhangpeng97@ihep.ac.cn
}
\address{\textsuperscript{\textit b}Nanjing University, Nanjing 210093, People’s Republic of China \\
jins@ihep.ac.cn}

\maketitle


\begin{abstract}
Radiative decays of the $J/\psi$ particle are of gluon-rich environment, providing an ideal place for hunting glueballs.
The $X(2370)$ particle was first discovered in $J/\psi\to \gamma \pi^+\pi^-\eta^{\prime}$ process in 2011 with the BESIII experiment at BEPCII Collider, and later it was confirmed in $J/\psi\to\gamma K\bar{K}\eta^{\prime}$ decays. In 2024, with a sample of 10 billion $J/\psi$ events collected at the BESIII detector, the spin-parity of the $X(2370)$ was determined to be $0^{-+}$ for the first time in the partial wave analysis of $J/\psi\to \gamma K^{0}_{S}K^{0}_{S} \eta^{\prime}$ process.
Recently, new decay modes of $X(2370)\to K^{0}_{S}K^{0}_{S}\pi^0$, $\pi^0\pi^0\eta$ and $a^0\pi^0$ were observed.
The mass, spin-parity quantum numbers, production and decay properties of the $X(2370)$ particle are consistent with the features of the lightest pseudoscalar glueball.
\end{abstract}


\ccode{PACS numbers: 03.65.$-$w, 04.62.+v}

\section{Introduction}

\begin{figure}[htbp]
    \centering
    \includegraphics[width=0.6\textwidth]{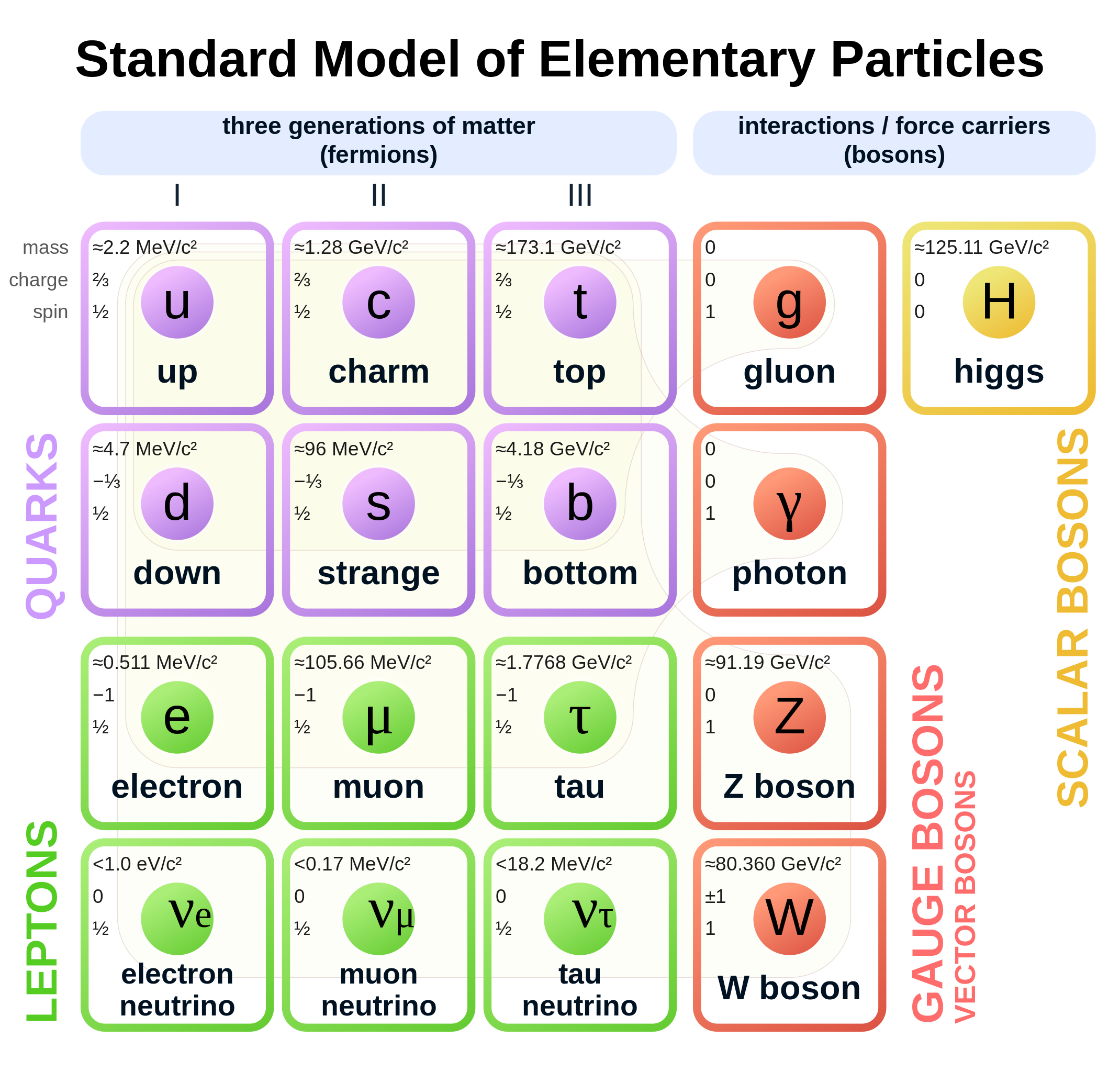}
    \caption{Elementary particles of the Standard Model. Taken from Wikimedia Commons.}
\label{fig:SM}
\end{figure}

In the particle physics, the Standard Model (SM)~\cite{Weinberg:1967tq, Glashow:1961tr, Salam:1968rm, tHooft:1972fi,Fritzsch:1973pi} successfully describes the elementary particles as well as the electroweak and strong interactions in the world (Figure~\ref{fig:SM}).

In the SM, there are leptons and quarks, which are all fermions, to form normal matter. The electroweak and strong interactions can be well described by the gauge field theory, and these interactions are mediated by the gauge bosons, including photons, W and Z bosons to mediate electroweak interactions and also gluons to mediate strong interactions.
The $W/Z$ bosons, leptons and quarks obtain their masses via interactions with the Higgs boson in the SM. The theory for the strong interaction part in the SM is called as Quantum Chromodynamics (QCD).

In the naive quark model, hadrons consist of two or three quarks. However, QCD allows new forms of hadrons, including multi-quark states which contain more than three quarks, hybrids which contain both quarks and gluons and glueballs which contain only gluons. Among them, glueballs are of special importance, since they are unique particles formed by pure gauge bosons due to non-Abel gauge self-interactions of gluons. The existence of glueballs would be a direct test of the  non-abel interactions of QCD theory.

\section{Glueball masses in Lattice QCD}
\label{sec:glueball_masses_in_lattice_qcd}

\begin{figure}[htbp]
    \centering
    \scalebox{0.55}{\input{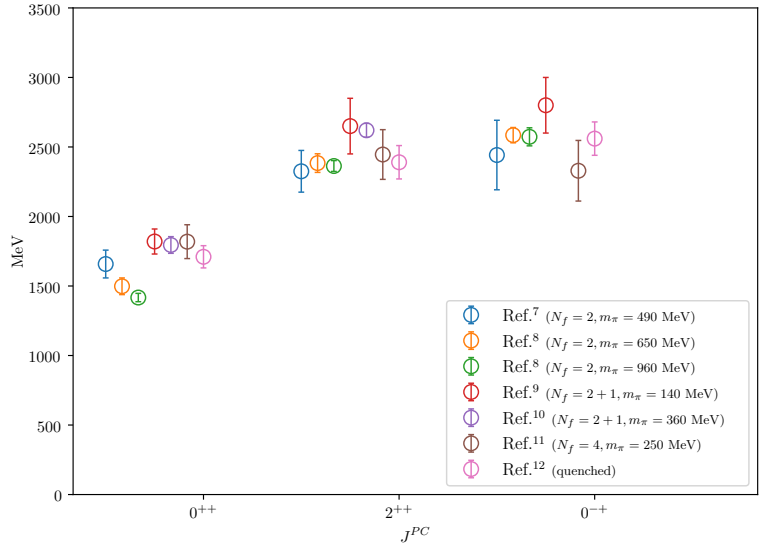}}
  \caption{A summary of estimates of the unquenched glueball
        spectrum. In light blue, the results from Ref.~\citen{Bali:2000vr},
    in light orange and green, the results from Ref.~\citen{Sun:2017ipk},
in red, the results in Ref.~\citen{Chen:2021dvn}, in purple the results
from Ref.~\citen{Gregory:2012hu}, in brown, the results from
Ref.~\citen{Athenodorou:2022nkb}, in cyan the quenched results
from Ref.~\citen{Chen:2005mg}. Taken from Ref.~\citen{Vadacchino:2023vnc}.
\label{fig:unquenched_lattice}}
\end{figure}

There are a couple of Lattice QCD (LQCD) calculations~\cite{Vadacchino:2023vnc} on the glueball masses.
Figure~\ref{fig:unquenched_lattice} in Ref.~\citen{Vadacchino:2023vnc}
summarizes various predictions on $0^{++}$, $2^{++}$ and $0^{-+}$ glueball masses from different LQCD  calculations.
Although there are still large uncertainties on the glueball mass predictions,
predictions from different LQCD groups, with either quenched LQCD or unquenched LQCD,  basically are consistent in mass ranges: for $0^{++}$ glueballs, the predicted mass range is $1.3-2.0~\GeV/c^{2}$;  $2^{++}$ glueball mass range is $2.2-2.8~\GeV/c^{2}$ and $0^{-+}$ glueball mass range is $2.3-3.0~\GeV/c^{2}$.

\section{Glueball production and decays}
\label{sec:glueball_production_decay}

\begin{figure}[htbp]
\centering

\subfloat{
  \includegraphics[width=0.45\textwidth]{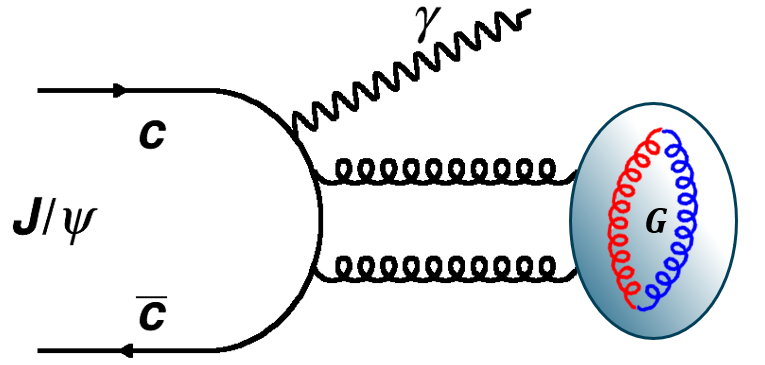}
  \label{fig:JpsiDecay_G}
\put(-160,75){(a)}
}
\subfloat{
  \includegraphics[width=0.45\textwidth]{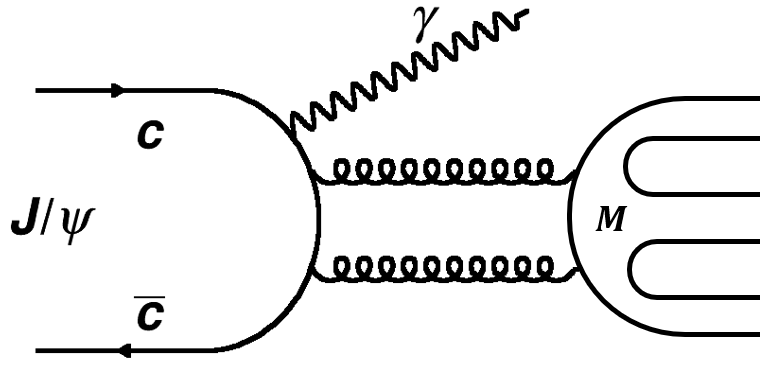}
  \label{fig:JpsiDecay_M}
\put(-160,75){(b)}
}

\caption{Glueball and normal meson productions in $J/\psi$ radiative decays.}
\label{fig:JpsiDecay}
\end{figure}

The $J/\psi$ radiative decays are of gluon-rich environment and they are believed to be an ideal place to search for glueballs
~\cite{Kopke:1988cs}.
The final hadron states in $J/\psi$ radiative decays are dominant by iso-spin 0 processes, and due to C-parity conservation, only positive C-parity is allowed for these final hadron states. So $J/\psi$ radiative decay processes are regarded as iso-spin and C-parity filters, which provides with clean background environments in performing glueball searchs.
This could be very different from many proton-antiproton collisions where all spin-parity and iso-spin processes are allowed and they can be mixed up and interference together.
Further more, from qualitative estimation of two diagrams of glueball and normal hadron production processes in $J/\psi$ radiative decays (Figure~\ref{fig:JpsiDecay}), the glueball production rate (proportional to $\alpha\alpha_{s}^{2}$) could be higher than that of a normal hadron (proportional to $\alpha\alpha_{s}^{4}$), so it could be even easier to search for glueballs in $J/\psi$ radiative decays than in other processes.

\begin{figure}[htbp]
\centering

\subfloat{
  \includegraphics[width=0.45\textwidth]{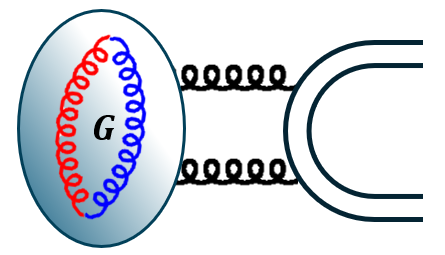}
  \label{fig:XDecay_glueball}
\put(-160, 90){(a)}
}
\subfloat{
  \includegraphics[width=0.45\textwidth]{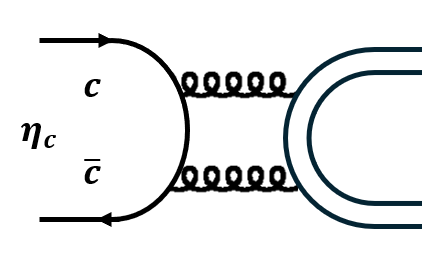}
  \label{fig:XDecay_etac}
\put(-160, 90){(b)}
}

\caption{$0^{-+}$ glueball and $\eta_{c}$ decays.}
\label{fig:XDecay}
\end{figure}

The glueball decays are expected to be flavor symmetric since they decay via gluons, which contain no flavors of quarks. Even though there are LQCD calculations on the glueball masses, the predictions on the glueball decay properties, including their decay patterns and decay branching fractions, are still missing due to theoretical difficulties. However, the glueball decays should be similar to the charmonium family decays since both of them decay via gluons~\cite{Chao:1995hd,Huang:1995td}. The plenty of charmoniun decay data from PDG \cite{pdg} could provide a good guidance in understanding the glueball decay properties.
For example, the pseudoscalar glueballs should have very similar decay properties to the $\eta_{c}$ as shown in Figure~\ref{fig:XDecay}, except that they may have different phase spaces decaying to the same final states due to their mass difference.
Direct comparison with $\eta_{c}$ decays is of crucial importance in the identification of a pseudoscalar glueball.

For $\eta_{c}$ decays, one of the largest decay modes is $\pi^+\pi^-\eta^{\prime}$, so the $J/\psi\to\gamma\pi^+\pi^-\eta^{\prime}$ process could be a good place to search for the $0^{-+}$ glueball.

$PPP$ decay modes (3 pseudoscalar meson decay modes such as $\pi^+\pi^-\eta$, $K^+K^-\eta$, $\pi^+\pi^-\eta^{\prime}$, $K^+K^-\eta^{\prime}$ and $K^+K^-\pi$) are believed to be golden decay modes for $0^{-+}$ gueball searches, because first they are decays in $S$-wave so that there are no suppression factors in the decays.
Secondly, the $PPP$ decays modes are forbidden for $0^{++}$ mesons and they are strongly suppressed for $2^{++}$ meson decays due to $D$-wave suppression. So $PPP$ modes are spin-parity filters for studying $0^{-+}$ mesons. For other decay modes:
1) $PP$ modes (2 pseudoscalar modes) are forbidden for $0^{-+}$ mesons.
2) $VV$ modes (2 vector meson decay modes), they are $P$-wave suppressed decays for $0^{-+}$ mesons and all spin-parity quantum numbers of mesons are allowed to decay to $VV$ modes, i.e., $VV$ modes are not spin-parity filters.
3) Baryon pair modes, all spin-parity quantum numbers of mesons are allowed to decay to baryon pairs and they are not spin-parity filters in the study of $0^{-+}$ mesons.

\section{Historical glueball candidates and difficulties}

Experimental observations of glueballs would be a crucial and direct test of QCD gauge theory, so glueballs have been searched for experimentally for more than 40 years since the beginning when QCD was established. There are a lot of candidates observed and studied at various $e^+e^-$, proton-antiproton and $\pi N$ interactions but none of them has been unambiguously identified as as glueball so far. The following will discuss about historical $0^{++}$, $2^{++}$ and $0^{-+}$ glueball candidates as well as the difficulties in glueball searches.

\subsection{$0^{++}$ glueball candidates: $f_{0}(1500)$ and $f_{0}(1710)$}

The LQCD predicted that the mass of scalar glueball lies in the range of $1.3-2.0~\GeV/c^{2}$. In this mass range, $f_{0}(1710)$ and $f_{0}(1500)$ were two of the historical candidates.

The $f_{J}(1710)$ was first discovered by the MARKII experiment in 1982 in $J/\psi\to\gamma K\bar{K}$ process~\cite{Etkin:1982se,Burke:1982am,Edwards:1981ex} and it was called as $\theta(1720)$. Its spin-parity was first determined as $2^{++}$ by MARKII experiment with only a simple fit to the angular distributions of $K\bar{K}$ system.
The measured spin-parity of $f_{J}(1710)$
was changed to $0^{++}$ by the BES experiment in 2003 with a full partial wave analysis (PWA) on $J/\psi\to\gamma K\bar{K}$ process~\cite{BES:2003iac} and it was recognized as a good candidate of scalar glueball.
Then BES and BESIII experiments performed comprehensive studies on $f_{0}(1710)$ with PWA in $J/\psi\to \gamma K\bar{K}$, $\gamma \pi\pi$, $\gamma \eta\eta$, $\gamma \eta\eta^{\prime}$~\cite{BES:2003iac,Ablikim:2006db,BESIII:2013qqz,BESIII:2022iwi}.
The high production rate of $f_{0}(1710)$ in $J/\psi$ radiative decay which is consistent with the LQCD calculation~\cite{Gui:2012gx} and suppression of $f_{0}(1710)\to \eta\eta^{\prime}$ decays further strongly supports that it has large glueball content~\cite{Brunner:2015oga}.

There are some difficulties for the glueball interpretation of $f_{0}(1710)$. In terms of decay properties, considering the similarities between glueball and charmonium families, the scalar glueball decays should be similar to $\chi_{c0}$ decays. However, $\chi_{c0}$ decays show a good symmetry between $\pi\pi$ and $K\bar{K}$ modes~\cite{pdg} just as decays via gluons,
while experimental results shows that $f_{0}(1710)$ favors to decay into $K\bar{K}$ mode. Also, the largest decay mode for $\chi_{c0}$ is $4\pi$ mode, but the $4\pi$ mode of $f_{0}(1710)$ has not been observed yet.

The $f_0(1500)$ was first observed in 1983~\cite{Gray:1983cw,Serpukhov-Brussels-AnnecyLAPP:1983xdr} in the $\pi\pi$ mode in the $\bar{p}N$ annihilations.
and it was also observed by the Crystal Barrel experiment in 1996 in the $4\pi$ mode in the $p\bar{p}$ collisions \cite{CrystalBarrel:1996wfh}.
It was believed as a good scalar glueball candidate since its mass is consistent with the scalar glueball mass from LQCD prediction while the $f_{0}(1710)$ was misidentified as $f_{2}(1720)$ at that time.
However, the BESIII results show that the measured branching fraction of $J/\psi \to \gamma f_{0}(1710)$
is one order of magnitude higher than that of the $f_{0}(1500)$ in final states of $\eta\eta$, $\pi^{0}\pi^{0}$, and $K^{0}_{S}K^{0}_{S}$~\cite{BESIII:2013qqz,BESIII:2015rug,BESIII:2018ubj},
which disfavors the glueball interpretation of the $f_{0}(1500)$.

The above difficulties might be due to possible strong mixing between the glueball and the $q\bar{q}$ components~\cite{Wang:2012qa,Lu:2013jj}. Then before the scalar glueball can be identified, we need to understand the mixing mechanism from the first principle, such as LQCD calculations, not only based on some phenomenological models.

\subsection{$2^{++}$ glueball candidates: $\xi(2230)$ and $f_{2}(2340)$}

The $\xi(2230)$ was first observed by MARKIII experiment in 1986 in $J/\psi\to \gamma K\bar{K}$ process~\cite{MARK-III:1985qfw}, and more evidences of $\xi(2230)$ were observed by BES experiment in 1996 in $J/\psi\to \gamma K\bar{K}$, $\gamma \pi\pi$, $\gamma p\bar{p}$ processes~\cite{BES:1996upd}.
It was believed as a tensor glueball candidate since its mass is consistent with the tensor glueball mass from LQCD prediction and it decays flavor symmetrically to $\pi\pi$ and $K\bar{K}$ modes~\cite{Huang:1995td}.
Unfortunately, the existence of $\xi(2230)$ was not confirmed in the later
higher statistical data from BES and BESIII.

There are several $f_{2}$ mesons in the mass range around $2.3~\GeV/c^{2}$~\cite{pdg}, among them, the $f_{2}(2340)$ has the largest production rate in $J/\psi$ radiative decays based on BESIII results~\cite{BESIII:2016qzq},
so it was regarded as a good candidate of tensor glueball.
However, in the data analysis experimentally , especially with PWA, the large number of $f_{2}$ mesons and their large overlaps each other in the mass distributions due to their wide widths make it difficult to obtain robust results on the $f_{2}(2340)$ decay branching fractions in each decay mode, then it is not easy to perform reliable comparisons between $f_{2}(2340)$ decays and the $\chi_{c2}$ decays.

\subsection{$0^{-+}$ glueball candidates: $\eta(1405)$}

Historically, the $\eta(1440)$\textemdash a pseudoscalar state around $1400~\MeV/c^{2}$ \textemdash was first observed in $p\bar{p}$ annihilation at rest into $\eta(1440)\pi^+\pi^-$ with $\eta(1440)\to\pi^+\pi^-\eta$ and $K\bar{K}\pi$~\cite{Baillon:1967zz}.
However, subsequent experiments indicate that the $\eta(1440)$ is likely to be the result of contributions from two distinct pseudoscalar states in this mass region: $\eta(1405)$ and $\eta(1475)$~\cite{Rath:1989rt,E852:2001ote,MARK-III:1990wgk,DM2:1990cwz,OBELIX:1995zjg,OBELIX:1997yvk,OBELIX:1999gnb,OBELIX:2002eai,Bolton:1992kb}.

It was believed as a good glueball candidate since it is richly produced in $J/\psi$ radiative decays and there were no LQCD predictions on any glueball masses yet.
Now $\eta(1405)$ is no longer believed as a pseudoscalar glueball candidate since its mass is far from the LQCD prediction.

Until 2011, there were no good $0^{-+}$ glueball candidates in the mass range above $2.3~\GeV/c^{2}$.

\section{BESIII at BEPCII}

\begin{figure}[htbp]
\centering

\subfloat{
  \includegraphics[width=0.35\textwidth]{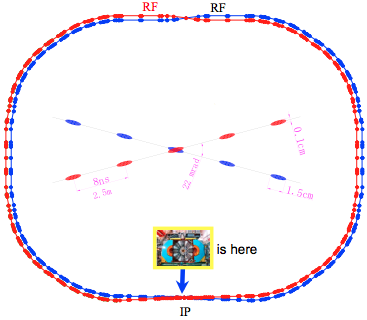}
  \label{fig:facility_BEPCII}
\put(-145, 95){(a)}
}
\subfloat{
  \includegraphics[width=0.5\textwidth]{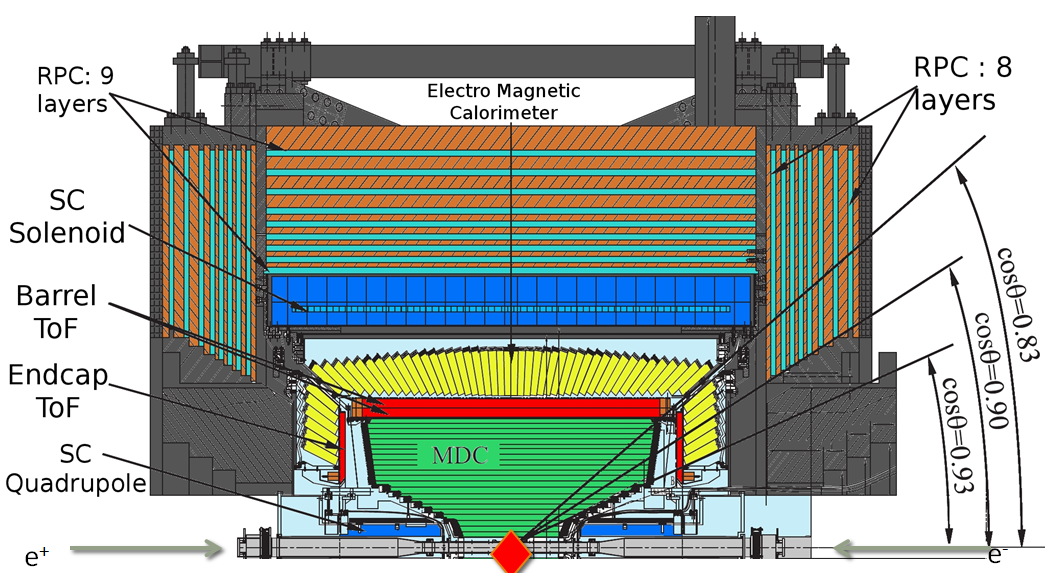}
  \label{fig:facility_BESIII}
\put(-185, 95){(b)}
}

\caption{Structure schematic of the BEPCII and BESIII.}
\label{fig:facility}
\end{figure}

The BESIII detector\cite{detector} records symmetric $e^+e^-$ collisions
provided by the BEPCII storage ring\cite{store_ring} in the center-of-mass energy range from 1.84 to 4.95~GeV,
with a peak luminosity of $1.1 \times 10^{33}\;\text{cm}^{-2}\text{s}^{-1}$ achieved at $\sqrt{s} = 3.773\;\text{GeV}$.
The structure schematic of the BEPCII and BESIII are shown in Figure~\ref{fig:facility}.
BESIII has collected large data samples in this energy region\cite{Ablikim:2019hff,EcmsMea,EventFilter}.
The cylindrical core of the BESIII detector covers 93\% of the full solid angle and consists of a helium-based
 multilayer drift chamber~(MDC), a time-of-flight system~(TOF), and a CsI(Tl) electromagnetic calorimeter~(EMC), which are all enclosed in a superconducting solenoidal magnet providing a 1.0~T magnetic field.
The magnetic field was 0.9~T in 2012, which affects 12.7\% of the total $J/\psi$ data.
The solenoid is supported by an octagonal flux-return yoke with resistive plate counter muon
identification modules interleaved with steel.
The charged-particle momentum resolution at $1~{\rm GeV}/c$ is $0.5\%$, and the
${\rm d}E/{\rm d}x$ resolution is $6\%$ for electrons from Bhabha scattering.
The EMC measures photon energies with a resolution of $2.5\%$ ($5\%$) at $1$~GeV in the barrel (end cap)
region. The time resolution in the plastic scintillator TOF barrel region is 68~ps, while that in the end cap region was 110~ps.
The end cap TOF system was upgraded in 2015 using multigap resistive plate chamber technology, providing a time resolution of 60~ps\cite{etof1,etof2,etof3}.

Currently, the total number of $J/\psi$ events collected by BESIII detector has reached about 10 billion\cite{number},
making it the largest $J/\psi$ sample in the world and providing a good opportunity to search for glueballs.

\section{Discovery of $X(2370)$}

The $X(1835)$ was the first new particle discovered by the BES experiment at BEPC collider in $J/\psi\to \gamma \pi^+\pi^-\eta^{\prime}$ process in 2005~\cite{BES:2005ega}, which was likely to be a $p\bar{p}$ bound state. So for the BESIII experiment, it was the first physics goal to confirm the $X(1835)$ in $J/\psi\to \gamma \pi^+\pi^-\eta^{\prime}$ in the new data with much higher statistics and better detector.
In the meantime, as we discussed in Section~\ref{sec:glueball_production_decay}, the $J/\psi\to \gamma \pi^+\pi^-\eta^{\prime}$ channel is a good place to search for the $0^{-+}$ glueball.

In 2011, with a sample of $(225.2 \pm 2.8) \times 10^6$ $J/\psi$ events at BESIII experiment, the analysis of $J/\psi\to \gamma \pi^+\pi^-\eta^{\prime}$ process was performed~\cite{2370_observed_2011}.
Due to the good momentum resolution and photon energy resolution of the BESIII detector, the $\eta^{\prime}$ signals can be well reconstructed and selected in both $\eta^{\prime}\to \pi^+\pi^-\eta$ and $\eta^{\prime}\to \gamma \pi^+\pi^-$ modes.

\begin{figure}[htbp]
    \centering
    \includegraphics[width=0.5\textwidth]{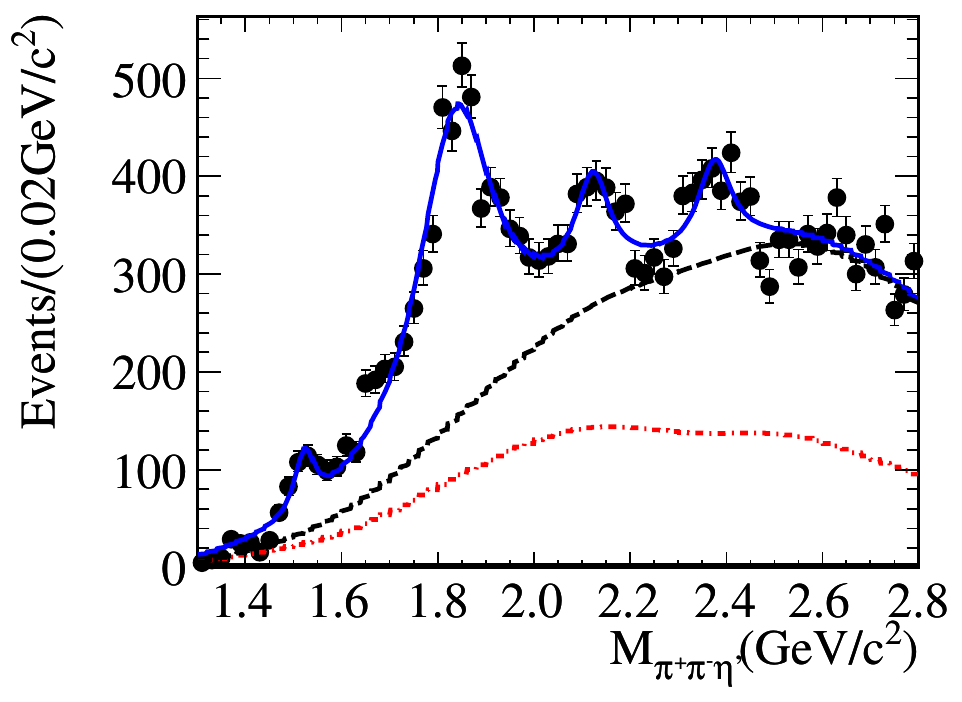}
    \caption{$X(2370)$ observation in $J/\psi \to \gamma \pi^+\pi^-\eta^{\prime}$ channel: The black dots are selected events from $\eta^{\prime}\to\gamma \pi^+\pi^-$ and $\eta^{\prime}\to \pi^+\pi^-\eta$ decay modes.
    The blue line represent the $\pi^+\pi^-\eta^\prime$ invariant mass spectrum fit result.
The dash-dot line represent contributions of non-$\eta^\prime$ events and the $\pi^0\pi^+\pi^-\eta^\prime$ background.
The black dash line represent contributions of the total background  and non-resonant $\pi^+\pi^-\eta^\prime$ process.
    Taken from Ref.\citen{2370_observed_2011}.
}
\label{fig:data_pipietap}
\end{figure}

In the $\pi^+\pi^-\eta^{\prime}$ mass spectrum of selected events shown in Figure~\ref{fig:data_pipietap},
besides the confirmation of the $X(1835)$, two new resonances, $X(2120)$ and $X(2370)$ were observed.
The mass, widths and statistical significances of the $X(1835)$, $X(2120)$ and $X(2370)$ are listed in Table~\ref{tab:gpipietap2011}.

\begin{table}[htbp]
\centering
 \caption{
The measured mass, widths and statistical significances of resonances observed in $J/\psi \to \gamma\pi^+\pi^-\eta^{\prime}$ channel.
}

\resizebox{\columnwidth}{!}{

\begin{tabular}{lccc}

\hline\hline
Resonance    & Mass ($\MeV/c^{2}$)            & Width (MeV)  & Significance \\
\hline

$X(1835)$ & $1836.5\pm3.0({\rm stat})^{+5.6}_{-2.1}({\rm syst})$ & $190\pm 9({\rm stat})^{+38}_{-36}({\rm syst})$ & $>20\sigma$ \\
$X(2120)$ & $2122.4\pm6.7({\rm stat})^{+4.7}_{-2.7}({\rm syst})$ & $83\pm 16({\rm stat})^{+31}_{-11}({\rm syst})$ & $7.2\sigma$ \\
$X(2370)$ & $2376.3\pm8.7({\rm stat})^{+3.2}_{-4.3}({\rm syst})$ & $83\pm 17({\rm stat})^{+44}_{-6}({\rm syst})$ & $6.4\sigma$ \\

\hline\hline
\end{tabular}
}

\label{tab:gpipietap2011}
\end{table}

In 2020, with a sample of $(1312.5\pm0.14)\times10^9$ $J/\psi$ events at the BESIII experiment, the combined analysis on $J/\psi \to \gamma K^+K^-\eta^{\prime}$ and $J/\psi \to \gamma K^{0}_{S}K^{0}_{S}\eta^{\prime}$, with the two $\eta^{\prime}$ primary decay modes of $\gamma \pi^+\pi^-$ and $\pi^+\pi^-\eta$,  was performed ~\cite{2370_observed_2020}. In this combined analysis, the $X(2370)$ was confirmed and a new decay mode of $X(2370)\to K\bar{K}\eta^{\prime}$ was observed with a statistical significance of $8.3\sigma$. The corresponding mass and width of the $X(2370)$ in this decay mode were measured to be $M_{X(2370)}=2341.6\pm6.5({\rm stat})\pm5.7({\rm syst})~{\rm
MeV}/c^2$ and $\Gamma_{X(2370)}=117\pm10({\rm stat})\pm8({\rm syst})~{\rm MeV}$, respectively.

\begin{figure}[htbp]
    \centering
    \includegraphics[width=0.5\textwidth]{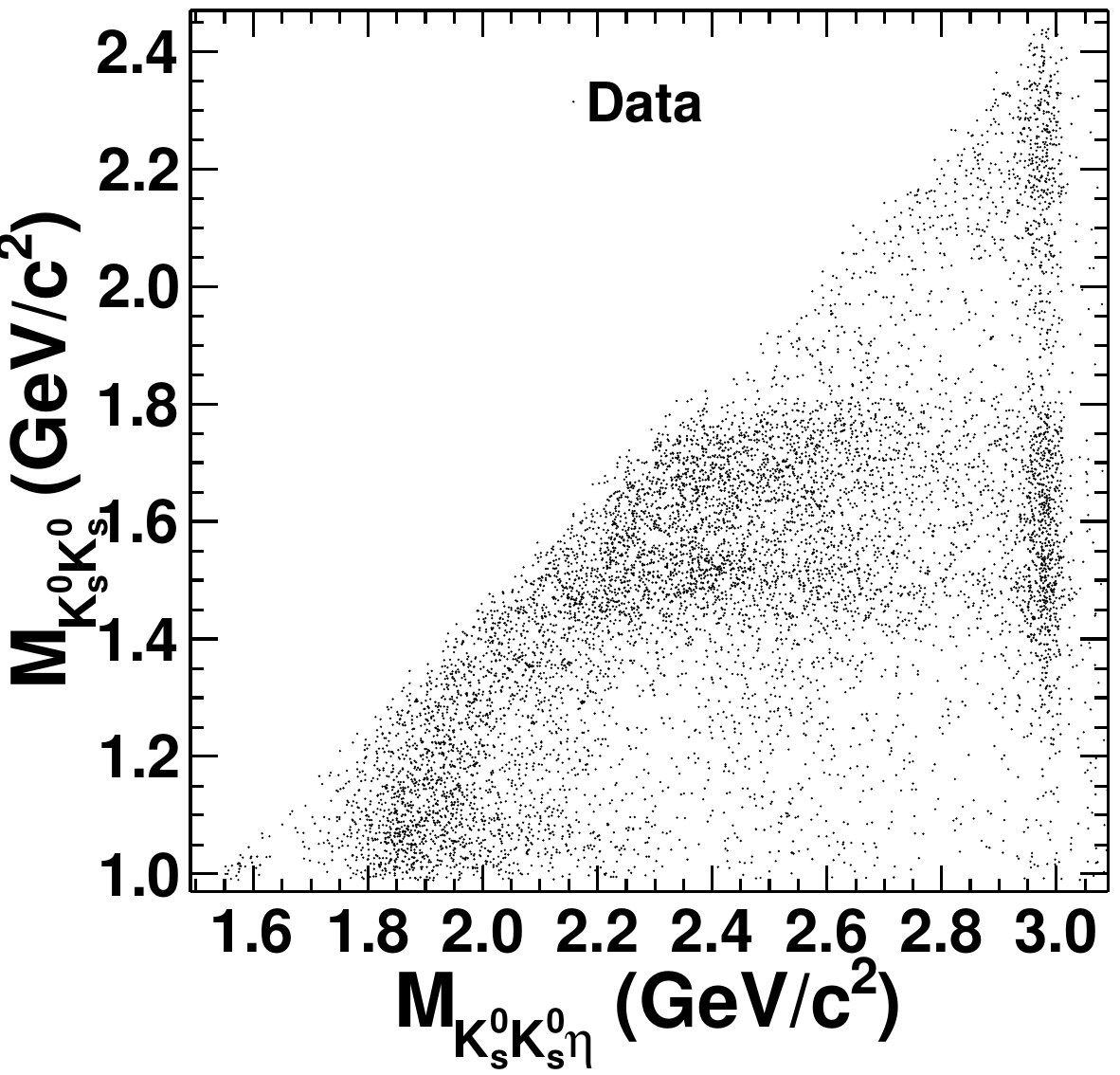}
    \caption{Scatter plot of $M_{K^{0}_{S}K^{0}_{S}}$ versus $M_{K^{0}_{S}K^{0}_{S}\eta}$ for selected events of $J/\psi \to \gamma K^{0}_{S}K^{0}_{S}\eta$ channel.
    Taken from Ref.\citen{1835jpc}.
}
\label{fig:data_ksks_kskseta}
\end{figure}

In the analysis of $J/\psi\to \gamma K^{0}_{S}K^{0}_{S}\eta$ channel, which firstly determined the spin-parity of the $X(1835)$ in 2015 at the BESIII experiment,
the scatter plot of $M_{K^{0}_{S}K^{0}_{S}}$ versus $M_{K^{0}_{S}K^{0}_{S}\eta}$ (Figure~\ref{fig:data_ksks_kskseta}) shows that in the mass band of the $f_{0}(1500)$ and the $f_{0}(1710)$,
there are clear evidences for both the $X(2370)$ and the $\eta_{c}$, while in the mass band of the $f_{0}(980)$, there are clear suppressions of both the $X(2370)$ and the $\eta_{c}$.

The $X(2370)$ was the first resonance observed in $J/\psi$ radiative decay with the mass consistent with the pseudoscalar glueball mass from the latest LQCD prediction.
Further more, the $X(2370)$ was observed in the golden decay modes of the $0^{-+}$ glueball decays,
which shows flavor symmetric behavior in $\pi^+\pi^-\eta^{\prime}$ and $K^+K^-\eta^{\prime}$ modes.
All those experimental results strongly indicate that the $X(2370)$ is a good candidate of pseudoscalar glueball.
Then determination of the spin-parity of the $X(2370)$ to see if it is really a $0^{-+}$ meson became extremely crucial.

\section{Spin-Parity determination of $X(2370)$}

The $X(2370)$ was discovered in the $J/\psi \to \gamma \pi^+\pi^-\eta^{\prime}$ process,
however, this channel is hardly to be used to perform a PWA for its spin-parity measurement
since it suffers a lot from the dominant background process of $J/\psi \to \pi^0 \pi^+\pi^-\eta^{\prime}$ due to possible misidentification of a $\pi^0$ as a photon.

Fortunately, the $X(2370)$ was confirmed in the $J/\psi \to \gamma K^{0}_{S}K^{0}_{S}\eta^{\prime}$ process,  in which there is no contamination background from $J/\psi \to \pi^0 K^{0}_{S}K^{0}_{S} \eta^{\prime}$ process since it is forbidden by the C parity conservation and the exchange symmetry of two $K^{0}_{S}$ mesons.
Therefore, the $J/\psi \to \gamma K^{0}_{S}K^{0}_{S}\eta^{\prime}$ decay provides a clean environment for the $J^{PC}$ measurement of the $X(2370)$ with minimal background modeling uncertainties. This enables us to perform a robust PWA in this process with a huge $J/\psi$ data sample collected at BEPCII and with a good performance of BESIII detector.

\begin{figure}[htbp]
\centering

\hspace{-3.2mm}
\subfloat{
  \includegraphics[width=0.45\textwidth]{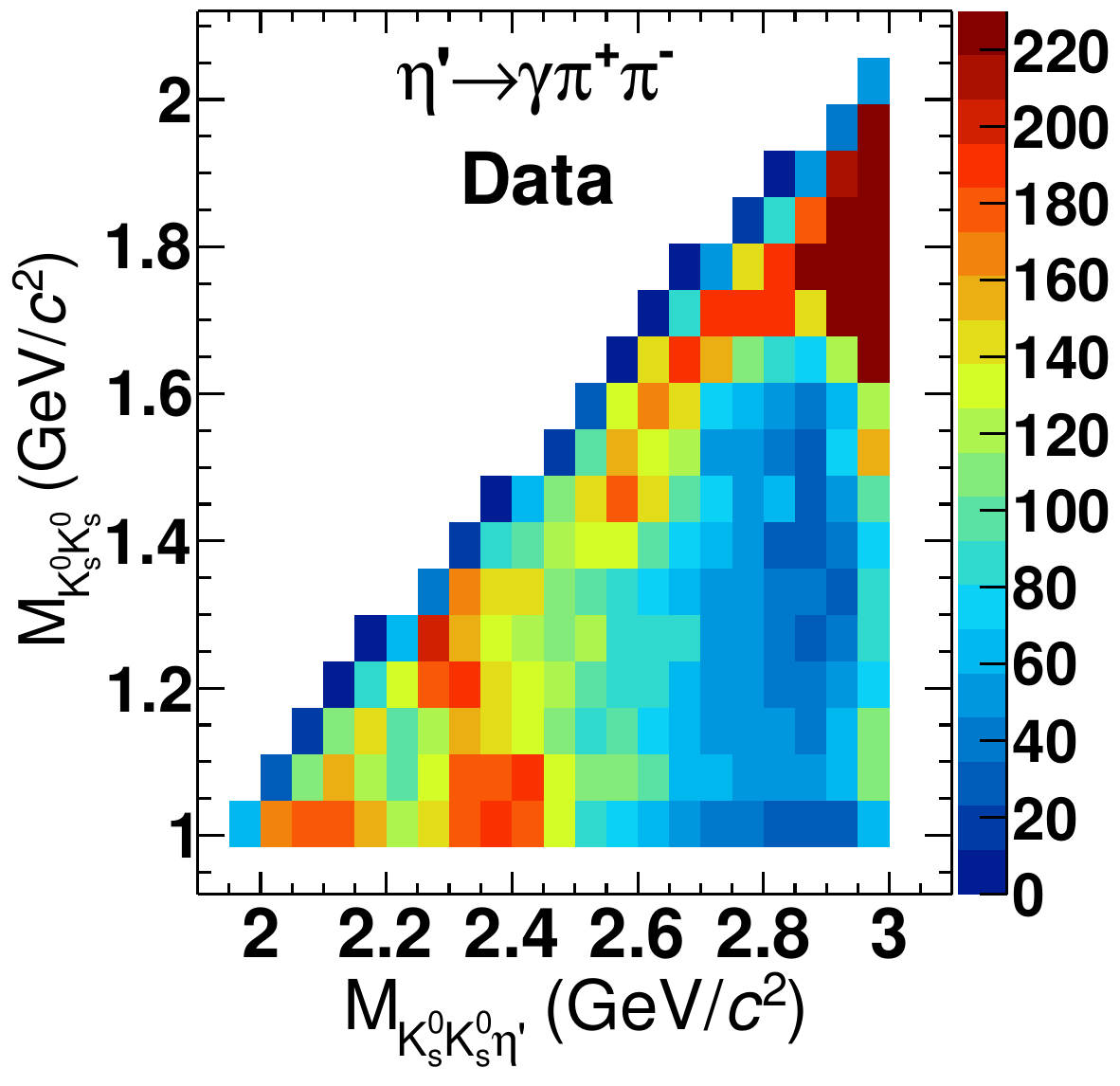}
  \label{fig:Subfigure1}
\put(-135,135){(a)}
}
\hspace{-3.2mm}
\subfloat{
  \includegraphics[width=0.45\textwidth]{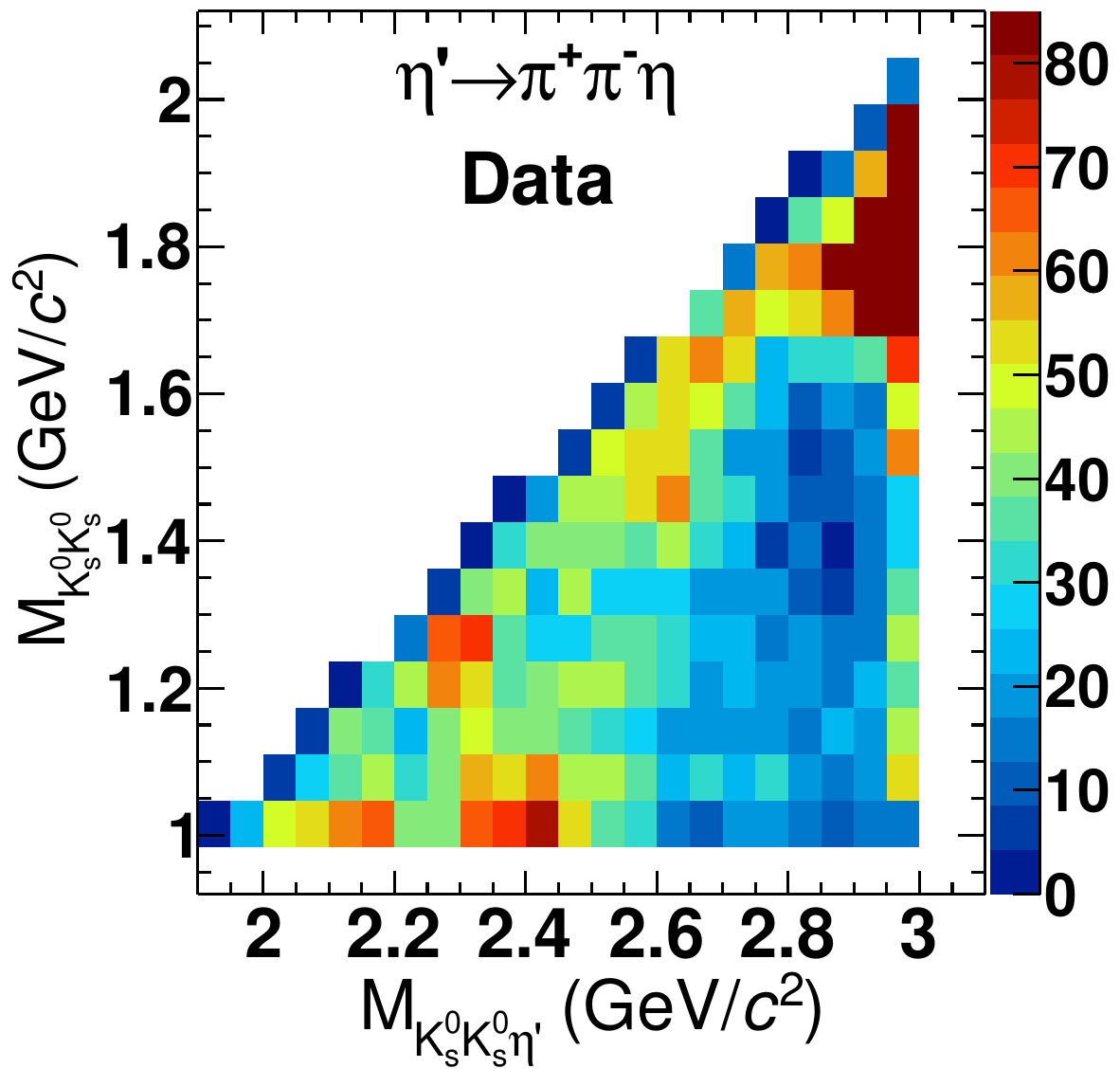}
  \label{fig:Subfigure2}
\put(-135,135){(b)}
}

\hspace{-3.2mm}
\subfloat{
  \includegraphics[width=0.45\textwidth]{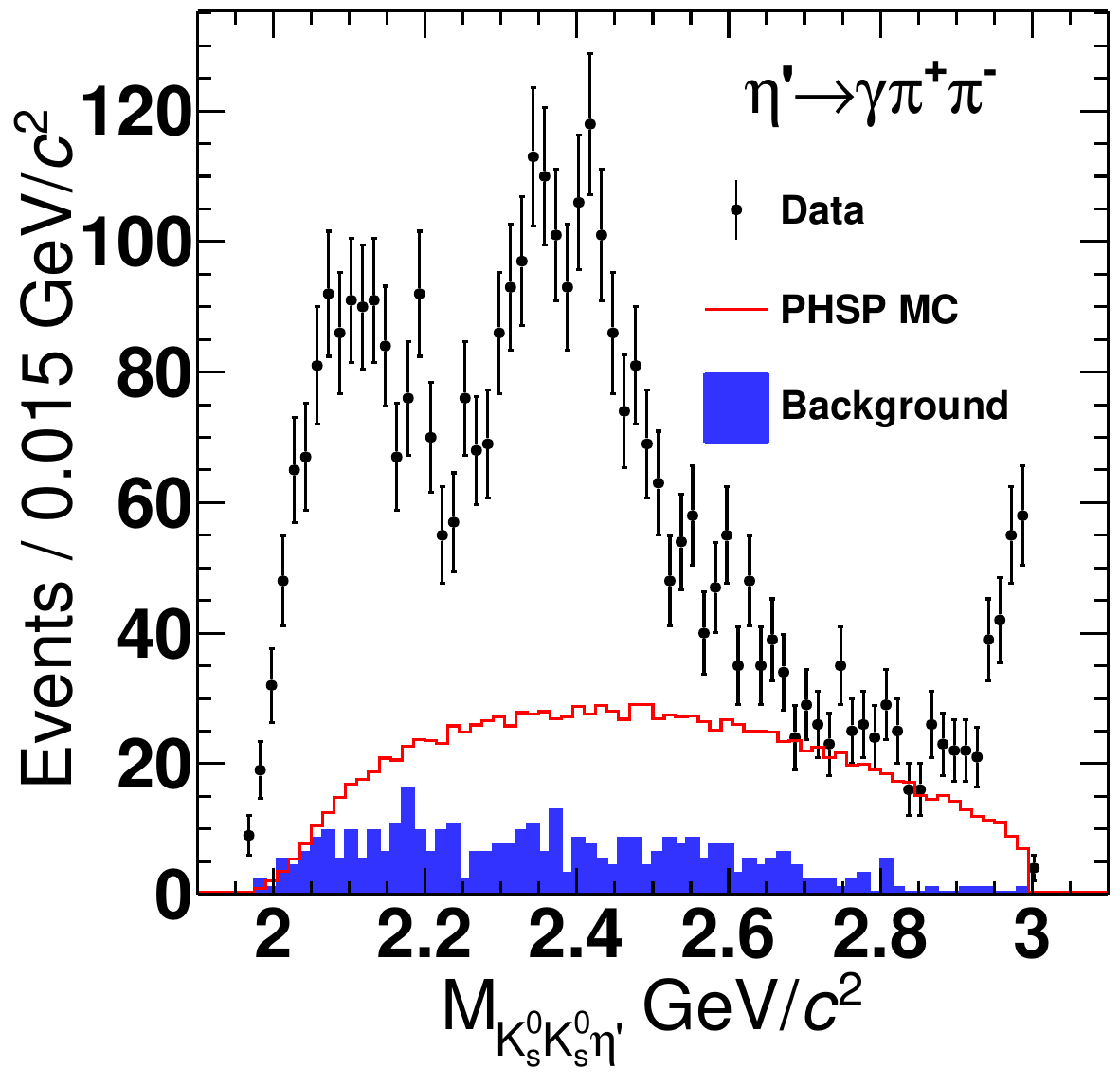}
  \label{fig:Subfigure3}
\put(-135,135){(c)}
}
\hspace{-3.2mm}
\subfloat{
  \includegraphics[width=0.45\textwidth]{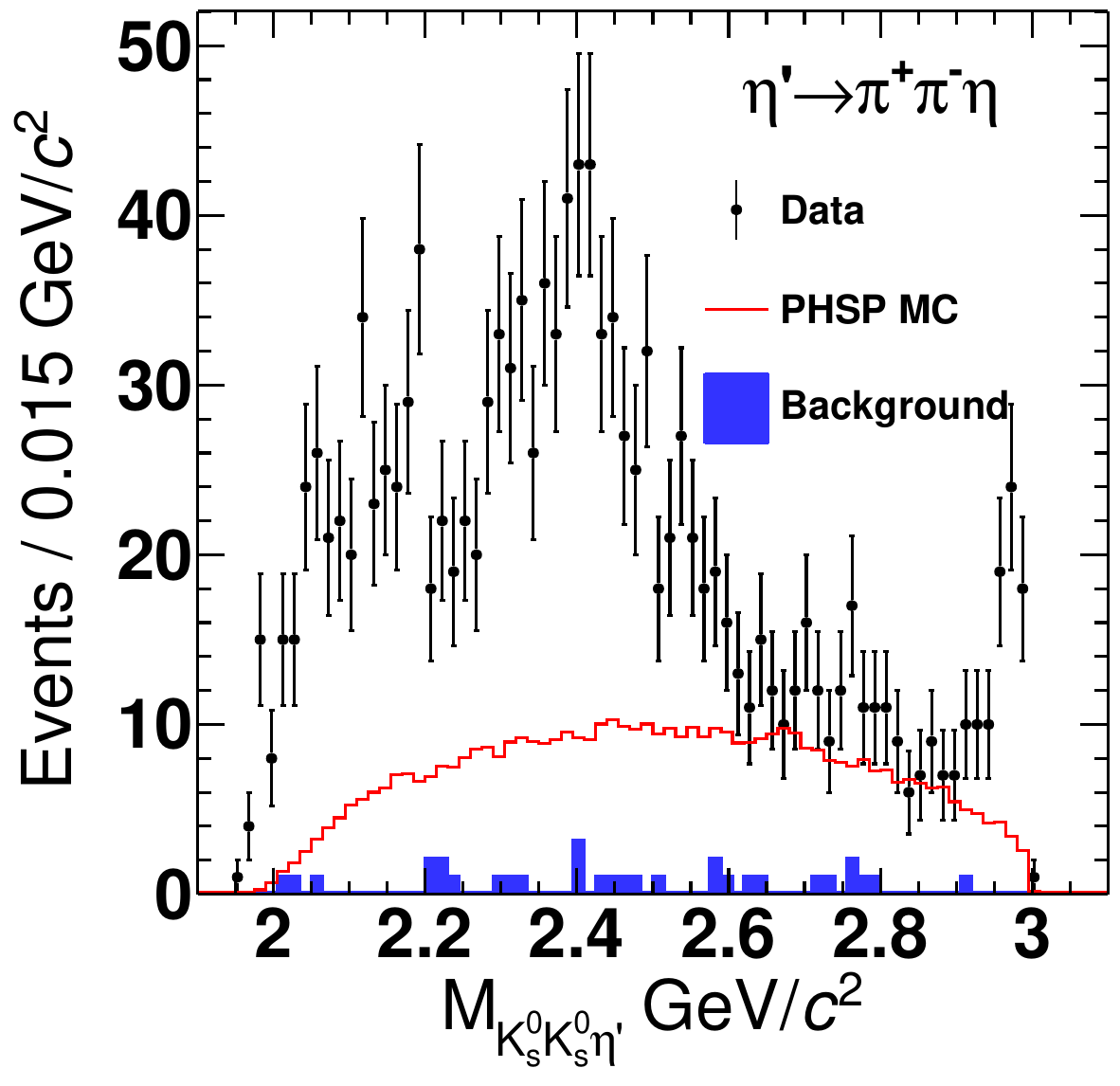}
  \label{fig:Subfigure4}
\put(-135,135){(d)}
}

\caption{
Invariant mass distributions of the selected events: (a) and (b) are the two-dimensional distributions of $M_{K_{S}^{0}K_{S}^{0}}$ versus $M_{K_{S}^{0}K_{S}^{0}\eta^{\prime}}$ for the $\eta^{\prime}\rightarrow\gamma\pi^{+}\pi^{-}$ and $\eta^{\prime}\rightarrow\pi^{+}\pi^{-}\eta$ channels, respectively.
(c) and (d) are the $K_{S}^{0}K_{S}^{0}\eta^{\prime}$ invariant mass distributions with the requirement $M_{K_{S}^{0}K_{S}^{0}}<1.1~\mathrm{GeV}/c^{2}$ for $\eta^{\prime}\rightarrow\gamma\pi^{+}\pi^{-}$ and $\eta^{\prime}\rightarrow\pi^{+}\pi^{-}\eta$ channels, respectively.
The dots with error bars are data. The shaded histograms are the non-$\eta^{\prime}$ backgrounds estimated by the $\eta^{\prime}$ sideband. The solid lines are phase space (PHSP) MC events with arbitrary normalization.
Taken from Ref.\citen{2370jpc}.
}

\label{fig:data_ksksetap}
\end{figure}

Recently, with a sample of $(10087 \pm 44) \times 10^6$ $J/\psi$ events at BESIII experiment, the selection of the $J/\psi \to \gamma K^{0}_{S}K^{0}_{S}\eta^{\prime}$ with the two $\eta^{\prime}$ decay modes of $\gamma\pi^+\pi^-$ and $\pi^+\pi^-\eta$ is performed. Due to the good reconstruction and resolution for the charge and neutral particle, both the $K^{0}_{S}$ and $\eta^{\prime}$ signals are well reconstructed with the mis-combination of $K^{0}_{S}$ reconstruction less than 0.1\%. The residual background contaminations from the non-$\eta^{\prime} $contribution are estimated to be 6.8\% and 1.8\% for the $\eta^{\prime} \to \gamma \pi^+\pi^-$ and $\eta^{\prime}\to\pi^+\pi^-\eta$ channels, respectively.
As shown in Figure~\ref{fig:data_ksksetap}, there is evident $f_{0}(980)$ signal in the low $K^{0}_{S}K^{0}_{S}$ mass threshold region, which shows clearly the connection between the $f_{0}(980)$ and the $X(2370)$. The $f_{0}(980)$ selection with the $M_{K^{0}_{S}K^{0}_{S}} <1.1 \GeV/c^{2}$ helps to reduce the complexities from additional intermediate processes.
In consequence, the $X(2370)$ becomes much more prominent in the $K^{0}_{S}K^{0}_{S}\eta^{\prime}$ mass spectrum. In addition, there is also a clear signal from the $\eta_{c}$.
In the PWA of this channel, the signal amplitudes are constructed with the covariant tensor formalism~\cite{zoubs2003} and parametrized as quasi-sequential two-body decays.
Due to the parity conservations,
the possible $J^{PC}$ of $K^{0}_{S}K^{0}_{S}\eta^{\prime}$ system are $0^{-+}$, $1^{++}$, $2^{++}$, $2^{-+}$, etc.

\begin{figure}[htbp]
\centering

\hspace{-4.0mm}
\subfloat{
  \includegraphics[width=0.45\textwidth]{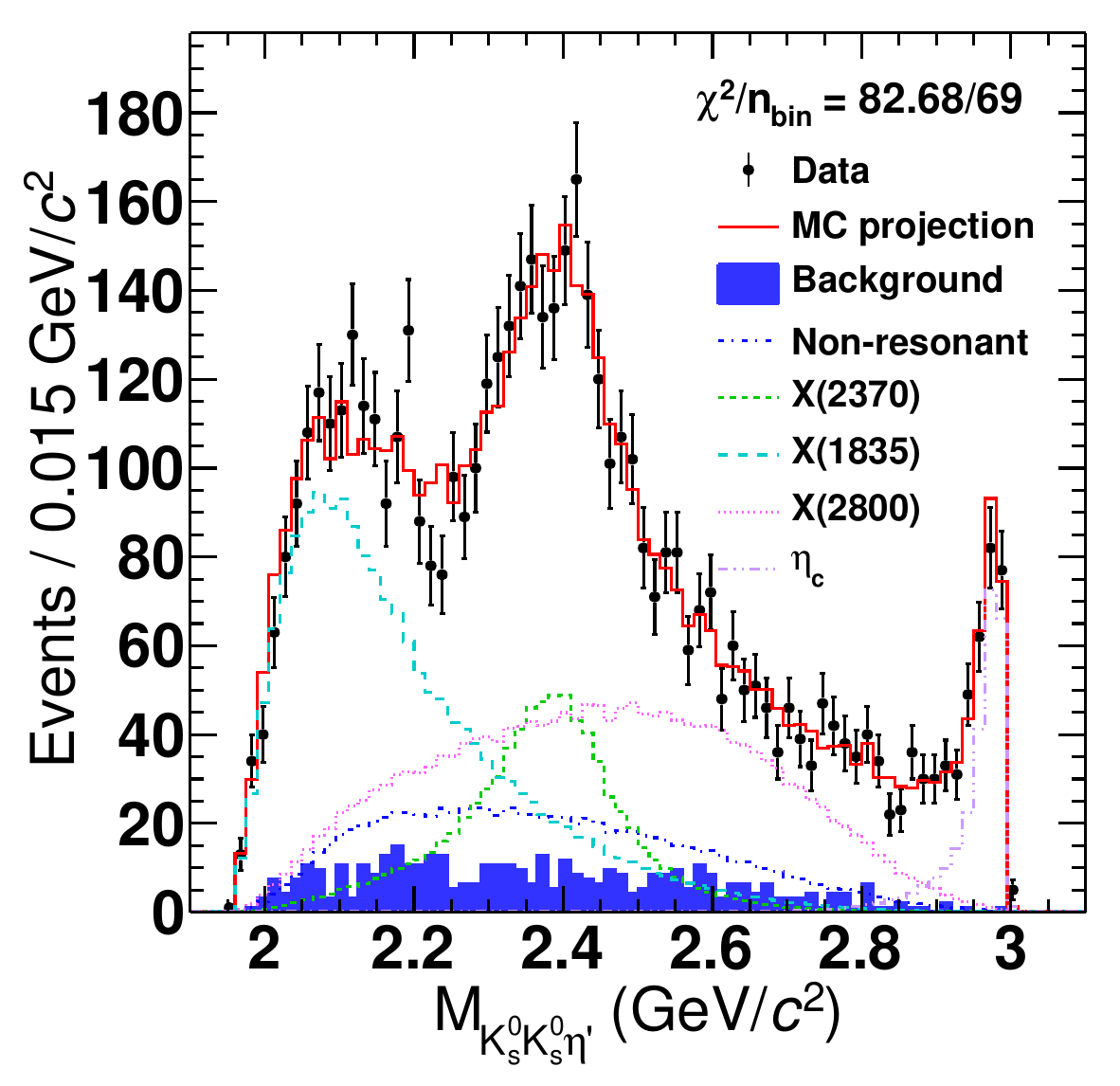}
  \label{fig:Subfigure11}
\put(-135,135){(a)}
}
\hspace{-4.0mm}
\subfloat{
  \includegraphics[width=0.45\textwidth]{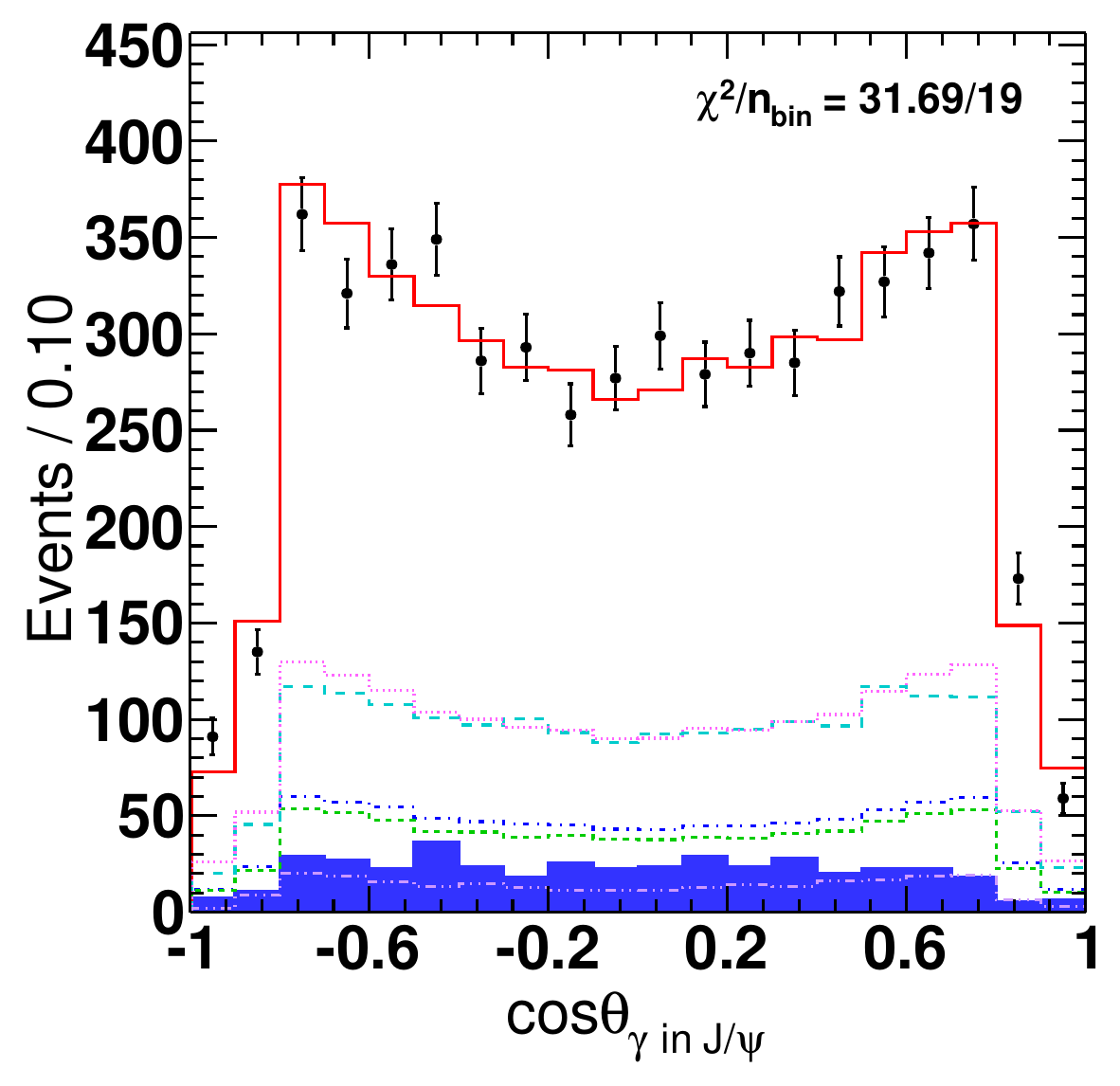}
  \label{fig:Subfigure12}
\put(-135,135){(b)}
}

 \caption{
  Comparisons between data (with two $\eta^{\prime}$ decay modes combined) and PWA fit projections: (a) is the invariant mass distributions of $K^{0}_{S}K^{0}_{S}\eta^{\prime}$, and (b) is the angular distributions of $\cos\theta$, where $\theta$ is the polar angle of $\gamma$ in the $J/\psi$ rest system. The dots with error bars are data. The solid red histograms are the PWA total projections. The shaded histograms are the non-$\eta^{\prime}$ backgrounds described by the $\eta^{\prime}$ sideband. The dash-dotted blue, short dashed green, long dashed cyan, dotted magenta and dash-dot-dotted violet show the contributions of the non-resonant contribution, $X(2370)$, $X(1835)$, $X(2800)$ and $\eta_{c}$, respectively.
  Taken from Ref.\citen{2370jpc}.
  }

  \label{fig:pwa_ksksetap}
\end{figure}

Figure~\ref{fig:pwa_ksksetap} shows the comparisons of the mass and angular distributions between data and PWA fit projects and there are good agreements.
The spin-parity investigation with complicated systematic studies shows that the spin-parity of the $X(2370)$ remains to be $0^{-+}$ with a significance greater than $10.1\sigma$ with respective to other alternative $J^{PC}$ hypotheses. It is the first time the $J^{PC}$ of the $X(2370)$ was determined to be $0^{-+}$.
The mass, width and production branching fraction of the $X(2370)$ are measured to be
$2395 \pm 11 ({\rm stat})^{+26}_{-94}({\rm syst})~\MeV/c^{2}$,
$188^{+18}_{-17}({\rm stat})^{+124}_{-33}({\rm syst})~\MeV$
and $\mathcal{B}[J/\psi\rightarrow\gamma X(2370)] \times \mathcal{B}[X(2370) \rightarrow f_{0}(980)\eta^{\prime}] \times \mathcal{B}[f_{0}(980) \rightarrow K^{0}_{S}K^{0}_{S}] =\left( 1.31 \pm 0.22 ({\rm stat})^{+2.85}_{-0.84}({\rm syst}) \right) \times 10^{-5}$, respectively.
The measured mass of the $X(2370)$ is in a good agreement with the mass prediction of the lightest pseudoscalar glueball from latest LQCD calculations~\cite{2370_prediction_glueball_2019}.

\section{Observation of new decay modes of the $X(2370)$}

Three $PPP$ decay modes, $\pi\pi\eta^{\prime}$, $K\bar{K}\eta^{\prime}$ and $K\bar{K}\eta$ modes, of $X(2370)$ have been observed at the BESIII experiment. Other two $PPP$ modes, $K\bar{K}\pi$ and $\pi\pi\eta$ modes, are also the two major decays modes of $\eta_{c}$,
so it is important to search for these two decay modes of the $X(2370)$.

Similar to the $J/\psi \to \gamma K_{S}^{0}K_{S}^{0} \eta^{\prime}$ decay, the $J/\psi \to \gamma K_{S}^{0}K_{S}^{0} \pi^0$ and $\gamma \pi^0\pi^0\eta$ decays are also almost background free since there are no possible dominant background contaminations from $J/\psi \to \pi^0 K_{S}^{0}K_{S}^{0} \pi^0$ and $\pi^0\pi^0\pi^0\eta$ due to C parity conservation and exchange symmetry.
Therefore, the analyses of $J/\psi \to \gamma K_{S}^{0}K_{S}^{0}\pi^0$ and $\gamma \pi^0\pi^0 \eta$ were performed to search for the $X(2370)$.

\subsection{Observation of the $X(2370)$ in $J/\psi\to\gamma K^{0}_{S}K^{0}_{S}\pi^{0}$ channel}

According to properties of final state of this channel,
each candidate event is required to have at least two positively charged tracks, at least two negatively charged tracks and at least three photons.
A four-constraint (4C) kinematic fit under the $J/\psi\to\gamma\gamma\gamma K^{0}_{S}K^{0}_{S}$ hypothesis is performed by enforcing energy-momentum conservation.
If there is more than one $\gamma\gamma\gamma K^{0}_{S}K^{0}_{S}$ combination, the one with the smallest $\chi^{2}_{\rm 4C}$ is chosen.
The reconstruction of the $K^{0}_{S}$ candidate follows the procedure described in Ref.~\citen{2370jpc}.

In order to reduce background,
a five-constraint (5C) kinematic fit is performed to further constrain the invariant mass of the two photons to $m_{\pi^{0}}$.
Among three $\gamma\gamma$ combinations, the one with the smallest $\chi^{2}_{\rm 5C}$ is chosen as $\pi^{0}$ candidate.

The $K^{0}_{S}$ and $\pi^{0}$ candidates are required to have reconstructed masses within approximately three standard deviations to their respective known masses.

The photon energy in selected combination is required to be greater than $100~\MeV$.
To suppressed background events containing a $\eta$ or multiple $\pi^{0}$,
events with $|M_{\gamma\gamma} - M_{\eta}| < 25~\MeV/c^{2}$ or $|M_{\gamma\gamma} - M_{\pi^{0}}| < 22~\MeV/c^{2}$ are rejected,
where the photon pairs are all possible combinations of the radiative photon and photons from $\pi^{0}$.
To suppressed background events containing $\omega$, events with $|M_{\gamma\pi^{0}} - M_{\omega}| < 40~\MeV/c^{2}$ are rejected.
With the above selection criteria, the miscombinations for the $K^{0}_{S}$ and $\pi^0$ signal are negligible.
The residual backgrounds from non-$K^{0}_{S}$ and non-$\pi^{0}$ contribution are 0.1\% and 0.2\%, respectively.

\begin{figure}[htbp]
    \centering
    \includegraphics[width=0.6\textwidth]{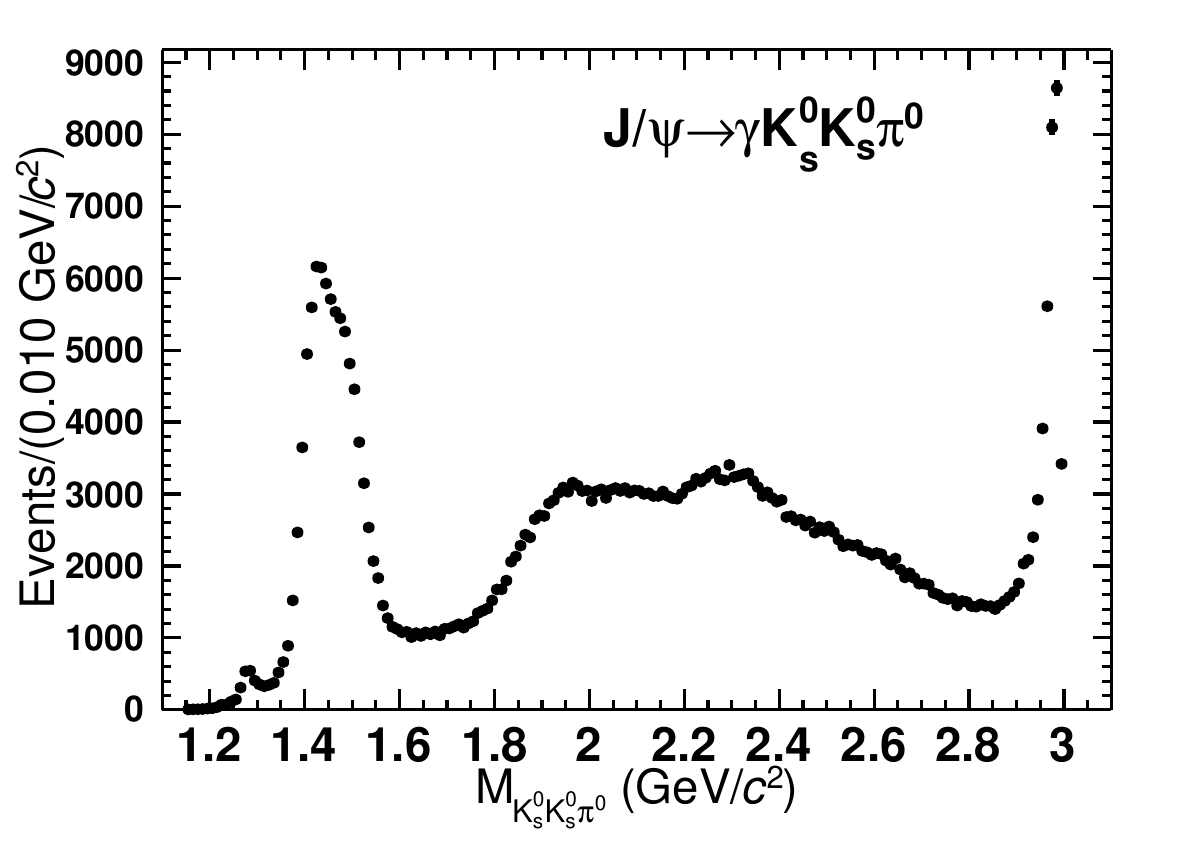}
            \put(-110, 90){\textit{\textsf{BESIII Preliminary}}}
    \caption{Invariant mass spectrum of $K^{0}_{S}K^{0}_{S}\pi^{0}$ for the selected data.}
\label{fig:data_kskspi0}
\end{figure}

Figure~\ref{fig:data_kskspi0} shows the mass distributions with the above selection criteria.
In the invariant mass spectrum of $K^{0}_{s}K^{0}_{s}\pi^{0}$,
there is a clear $X(2370)$ peak around $2.3~\GeV/c^{2}$,
along with the $\eta_{c}$ peak around $2.9~\GeV/c^{2}$.

\begin{figure}[htbp]
    \centering
    \includegraphics[width=0.6\textwidth]{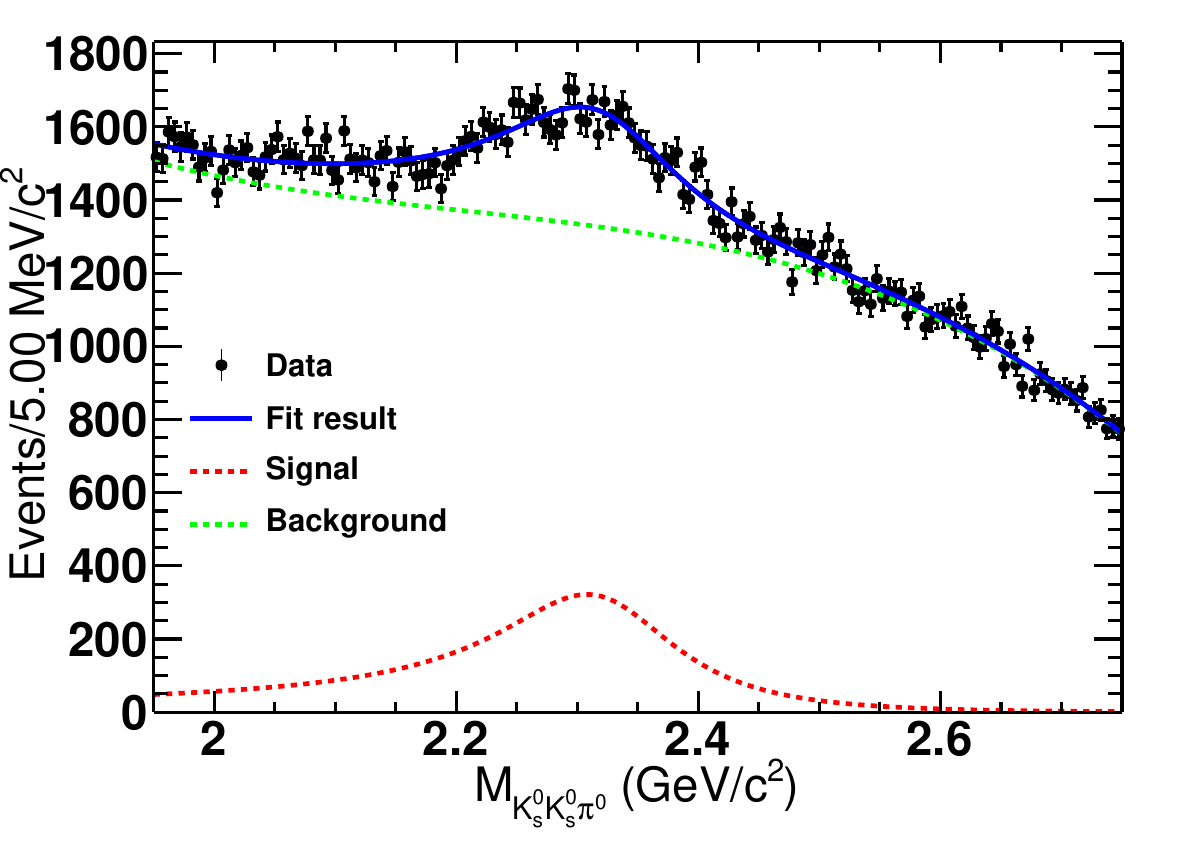}
        \put(-110, 70){\textit{\textsf{BESIII Preliminary}}}
    \caption{The fit result of the $K^{0}_{s}K^{0}_{s}\pi^{0}$ mass spectrum.
    The black points with error bars represent data, and the blue solid line is the total fit.
    The red dashed line and green dashed line describes the $X(2370)$ signal and continuum background, respectively.}
    \label{fig:fit_kskspi0}
\end{figure}

An unbinned maximum-likelihood fit is performed on the invariant mass spectrum of $K^{0}_{s}K^{0}_{s}\pi^{0}$ between 2.0 and $2.7~\GeV/c^{2}$,
 as shown in Figure~\ref{fig:fit_kskspi0}.
The $X(2370)$ signal is described by an efficiency-corrected BW function,
incorporating the phase space factor $E^{3}_{\gamma}$ for process $J/\psi \to \gamma 0^{-+} \to \gamma K^{0}_{s}K^{0}_{s}\pi^{0}$.

The continuum background arising from other processes with the $\gamma K^{0}_{s}K^{0}_{s}\pi^{0}$ final state are described with a third-order Chebyshev polynomial function.
Different systematic variations are performed,
including the consideration of interference between the signal and background functions,
as well as variations in the phase space factor, the background function and the fit range.
After evaluating with the consideration of all above systematic uncertainty variations,
The statistical significance of the $X(2370)$ is much greater than $5\sigma$.
The mass and width of the $X(2370)$ are determined to be
$M_{X(2370)} = 2321\pm4({\rm stat})\pm65({\rm syst})$ and $\Gamma_{X(2370)} = 182\pm 16({\rm stat})\pm59({\rm syst})$, respectively.

\subsection{Observation of the $X(2370)$ in $J/\psi\to\gamma \pi^{0}\pi^{0}\eta$ channel}

For the selection of $J/\psi\to\gamma \pi^{0}\pi^{0}\eta$ signature,
each candidate event is required to have zero charged tracks and at least seven photons.
Any photon detected in the barrel (end cap) portion of the EMC must have an energy of at least 25 (50) MeV.
To identify all pairs of two photons that may each originate from a $\pi^{0}$,
a one-constraint (1C) kinematic fit is performed by
constraining the invariant mass of the two photons to $m_{\pi^{0}}$.
Among all photons, at least two pairs are required to have $\chi^{2}_{\rm 1C}<10$, and these pairs are chosen as $\pi^{0}$ candidate.
Then,
a six-constraint (6C) kinematic fit under the $J/\psi\to\gamma\gamma\gamma \pi^{0}\pi^{0}$ hypothesis is performed by enforcing energy-momentum conservation and constraining the masses of two pairs of photons to $m_{\pi^{0}}$.
The combination with the smallest $\chi^{2}_{\rm 6C}$ is chosen if more than one combination is found.
Based on the photon combinations chosen by 6C kinematic fit,
To reduce miscombination, a seven-constraint (7C) kinematic fit is performed to further constrain the invariant mass of the two photons to $m_{\eta}$,
Among three $\gamma\gamma$ combinations, the one with the smallest $\chi^{2}_{\rm 7C}$ is chosen as $\eta$ candidate.

The $\eta$ candidate is required to have reconstructed masses within approximately three standard deviations to its known mass.

To suppress multi-photon background,
the value of $\chi^{2}_{4C}$ from the kinematic fit under the $J/\psi\to 7 \gamma$ hypothesis is required to be less than that from the kinematic fits under the  $J/\psi\to 8 \gamma$ or  $J/\psi\to 9 \gamma$ hypothesis.
To suppress background events containing misidentified $\eta$ or $\pi^{0}$,
events with any of the following requirements are rejected:
$| M_{\gamma_{\text rad}\gamma_{\eta/\pi^{0}}} - m_{\pi^{0}} | < 20~\MeV/c^2$,
$| M_{\gamma_{\text rad}\gamma_{\pi^{0}}} - m_{\eta} | < 30~\MeV/c^2$,
or $| M_{\gamma_{\text rad}\gamma_{\eta}} - m_{\eta} | < 50~\MeV/c^2$.
Here, $\gamma_{\text{rad}}$ refers to the radiative photon,
and the deviations correspond to $3\sigma$ for the respective known masses.
To suppressed background events containing $\omega$,  events with $|M_{\gamma\pi^{0}} - M_{\omega}| < 40~\MeV/c^{2}$ are rejected.

\begin{figure}[htbp]
\centering
\vspace{-2.0mm}

\subfloat{
  \includegraphics[width=0.5\textwidth]{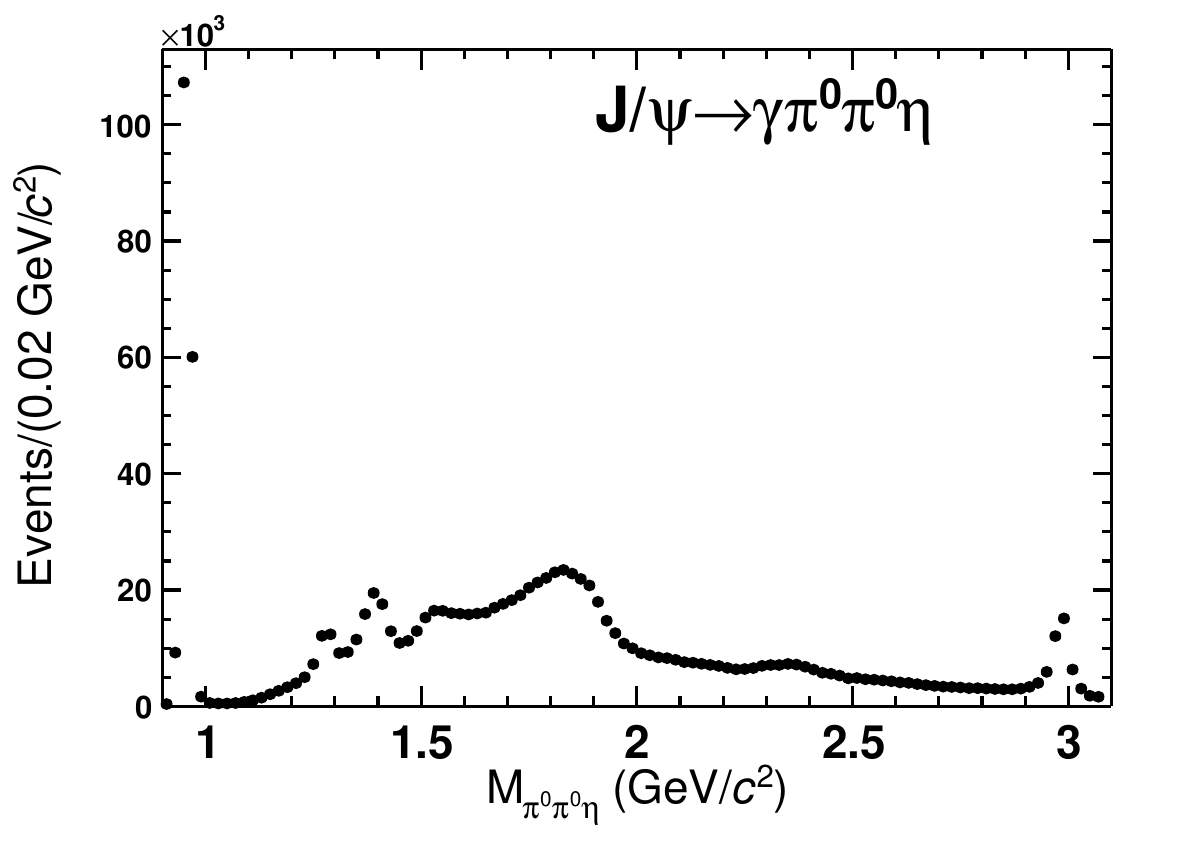}
  \label{fig:data_pi0pi0eta_1}
\put(-150,110){(a)}
        \put(-110, 90){\textit{\textsf{BESIII Preliminary}}}
}
\subfloat{
  \includegraphics[width=0.5\textwidth]{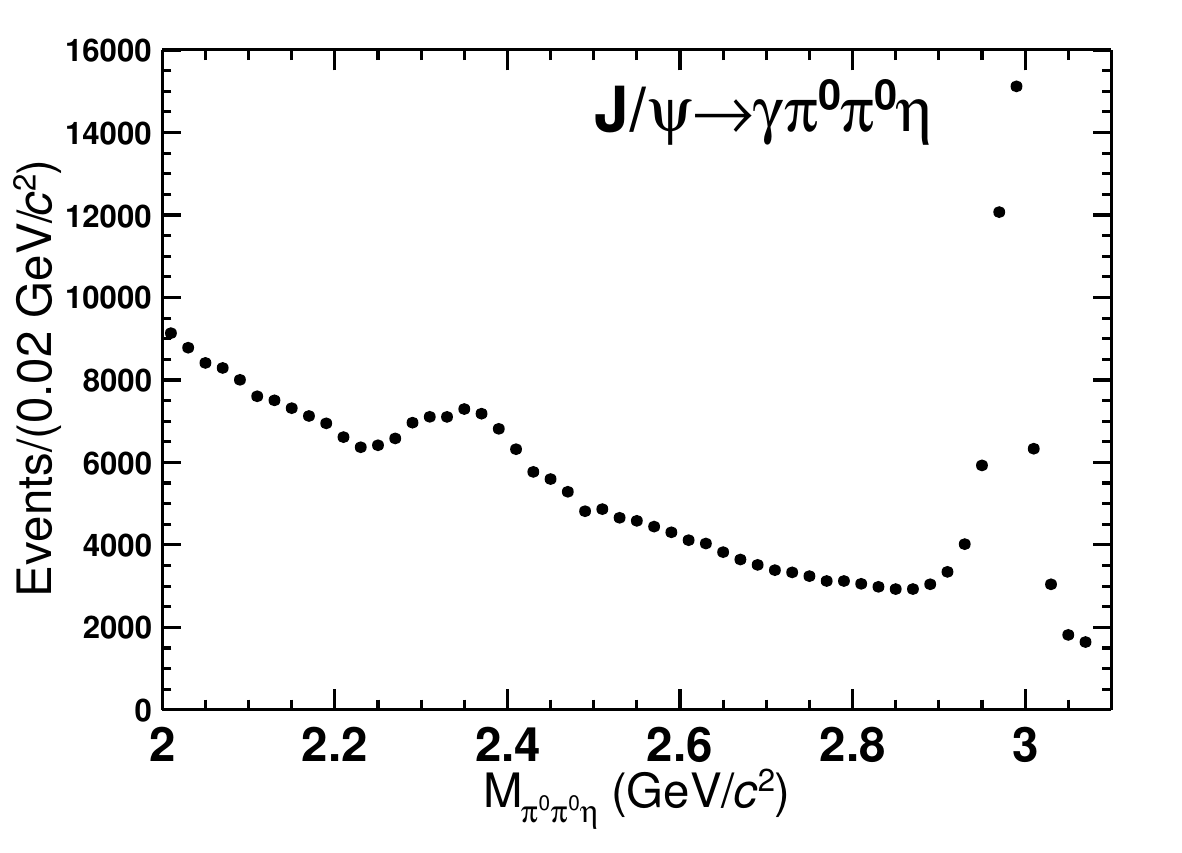}
  \label{fig:data_pi0pi0eta_2}
\put(-150,110){(b)}
        \put(-110, 90){\textit{\textsf{BESIII Preliminary}}}
}
\vspace{-3.0mm}

\caption{
Invariant mass distributions of the selected data: (a) and (b) show different ranges of $M_{\pi^{0}\pi^{0}\eta}$.
}

\label{fig:data_pi0pi0eta}
\end{figure}

Figure~\ref{fig:data_pi0pi0eta} shows the mass distributions with the above selection criteria.
In the invariant mass spectrum of $\pi^{0}\pi^{0}\eta$,
there is a clear $X(2370)$ peak around $2.3~\GeV/c^{2}$,
along with the $\eta_{c}$ peak around $2.9~\GeV/c^{2}$.

\begin{figure}[htbp]
    \centering
    \includegraphics[width=0.6\textwidth]{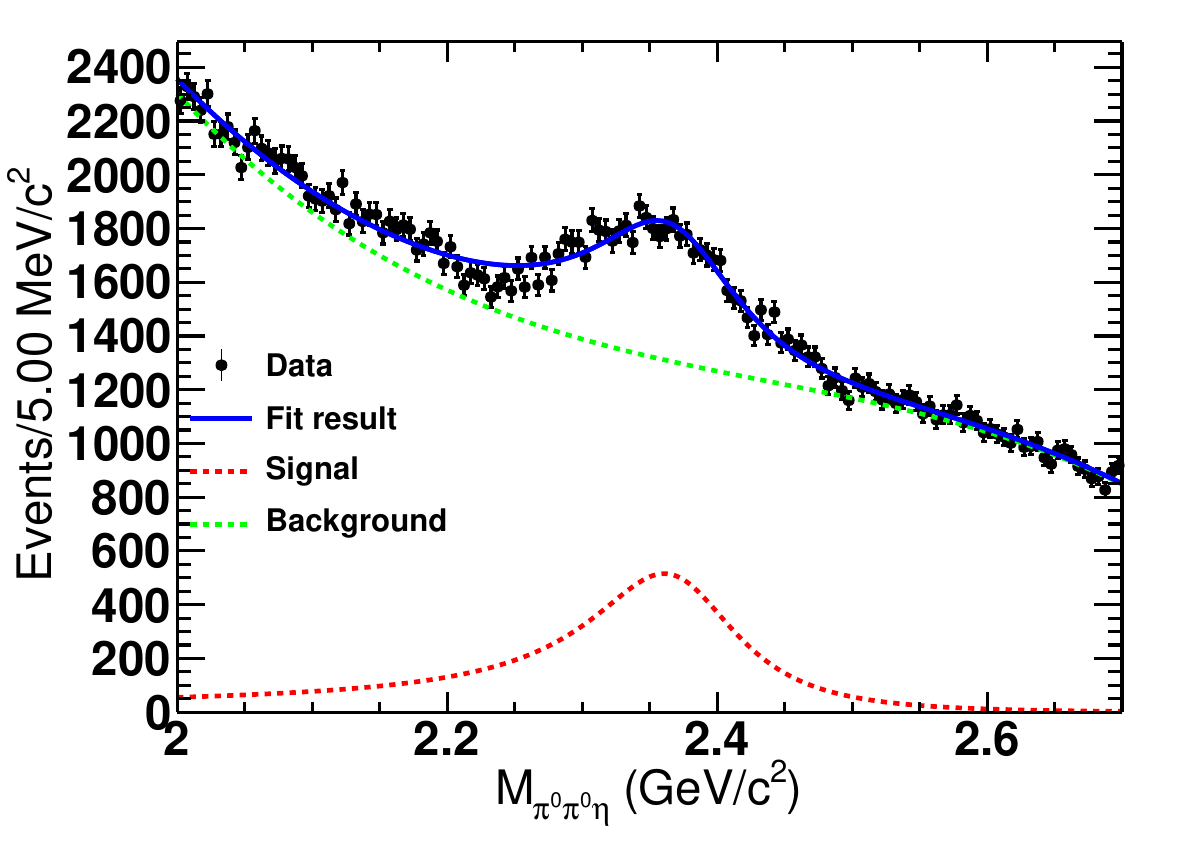}
        \put(-110, 60){\textit{\textsf{BESIII Preliminary}}}
    \caption{The Fit result of the $\pi^{0}\pi^{0}\eta$ mass spectrum.
    The black points with error bars represent data, and the blue solid line is the total fit.
    The red dashed line and green dashed line describes the $X(2370)$ signal and continuum background, respectively.}
\label{fig:fit_pi0pi0eta}
\end{figure}

An unbinned maximum-likelihood fit is performed to the invariant mass spectrum of $\pi^{0}\pi^{0}\eta$ between 2.0 and $2.7~\GeV/c^{2}$,
following the same fitting strategy used for the $J/\psi\to\gamma K^{0}_{S}K^{0}_{S}\pi^{0}$ channel,
 as shown in Figure~\ref{fig:fit_pi0pi0eta}.
The systematic uncertainties for this channel are evaluated, including the same sources considered in the analysis of $J/\psi\to\gamma K^{0}_{S}K^{0}_{S}\pi^{0}$ channel.
Considering all systematic uncertainties,
the statistical significance of the $X(2370)$ is much greater than $5\sigma$.
The mass and width of the $X(2370)$ are determined to be
$M_{X(2370)} = 2370\pm2 ({\rm stat})~\MeV/c^{2}$ and $\Gamma_{X(2370)} = 133\pm 8({\rm stat})~\MeV$, respectively.

\begin{figure}[htbp]
\centering
\vspace{-2.0mm}

\subfloat{
  \includegraphics[width=0.5\textwidth]{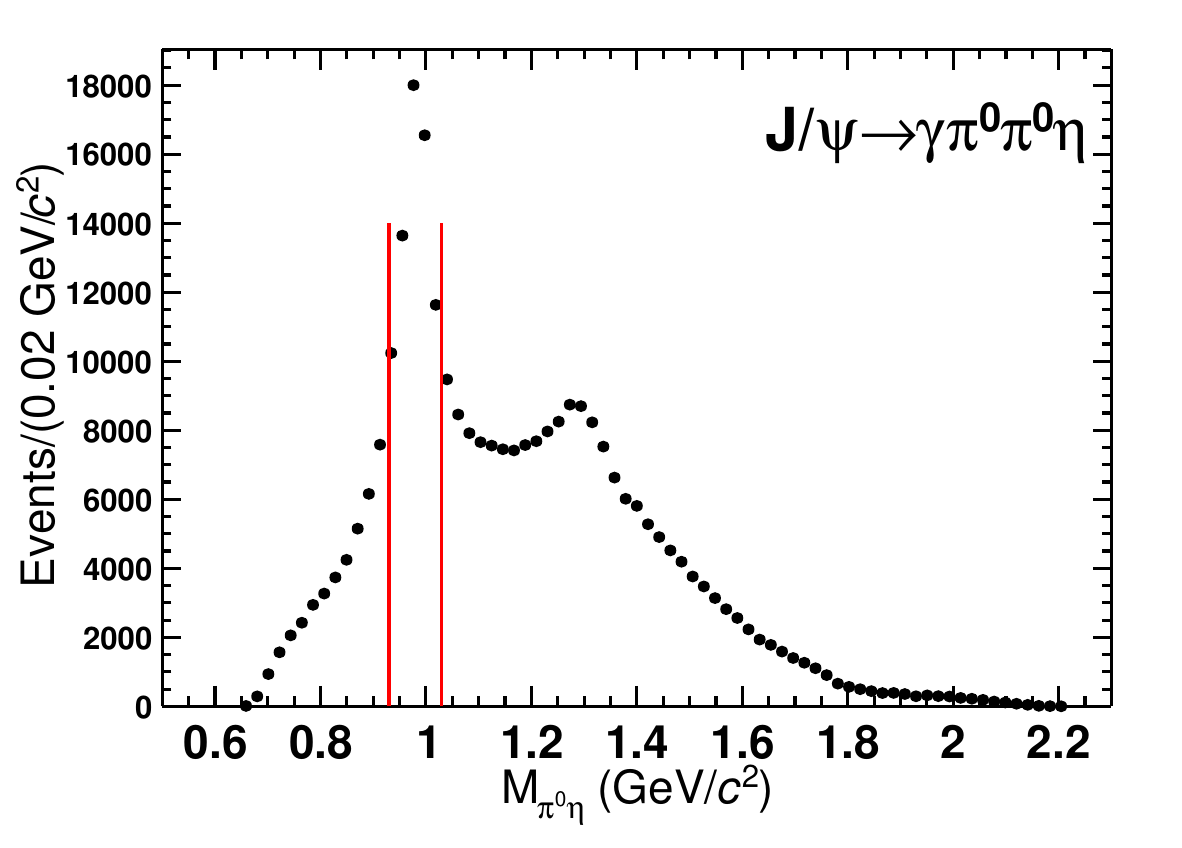}
  \label{fig:data_pi0pi0eta_a0_1}
\put(-150,110){(a)}
        \put(-110, 90){\textit{\textsf{BESIII Preliminary}}}
}
\subfloat{
  \includegraphics[width=0.5\textwidth]{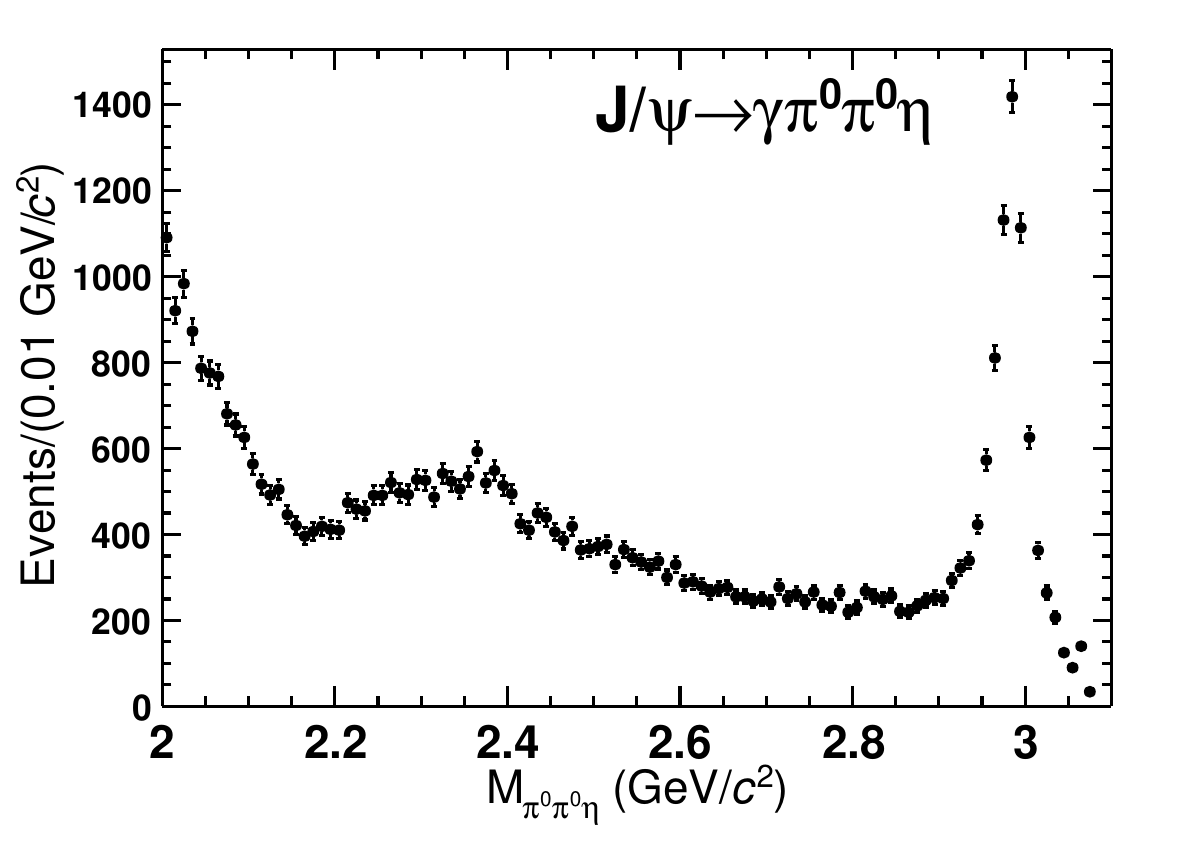}
  \label{fig:data_pi0pi0eta_a0_2}
\put(-150,110){(b)}
        \put(-110, 90){\textit{\textsf{BESIII Preliminary}}}
}
\vspace{-3.0mm}

\caption{
Invariant mass distributions of the selected data: (a) $M_{\pi^{0}\eta}$ and (b) $M_{\pi^{0}\pi^{0}\eta}$.
The red line in (a) indicate the required $a_{0}(980)$ region.
}

\label{fig:data_pi0pi0eta_a0}
\end{figure}

Based the selection criteria of $J/\psi\to\gamma \pi^{0}\pi^{0}\eta$ channel described above,
the region $|m_{\pi^{0}\eta}-0.98|<0.05~\GeV/c^2$ is required to select $a_{0}(980)$ signal,
where the $a_{0}(980)$ signal is constructed with $\eta$ and the $\pi^{0}$ with lower energy.
The event with $|M_{\pi^{0}\pi^{0}} - 1.5 |< 0.15~\GeV/c^2$
are further rejected to suppress the contribution from $X(2370)\to f_{0}(1500)\eta$.
The $\pi^{0}\pi^{0}\eta$ mass spectrum with above selection criteria is shown in Figure~\ref{fig:data_pi0pi0eta_a0}.

\begin{figure}[htbp]
    \centering
    \includegraphics[width=0.6\textwidth]{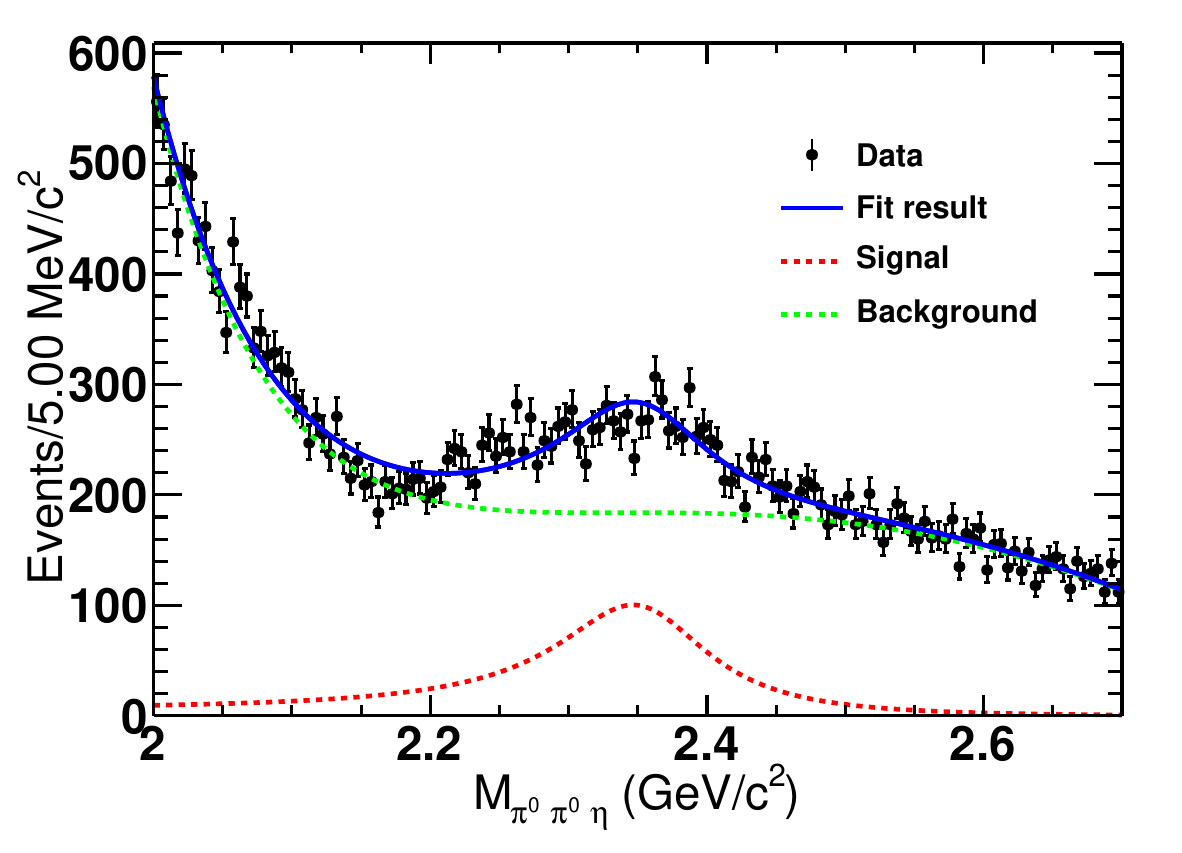}
            \put(-170, 120){\textit{\textsf{BESIII Preliminary}}}
    \caption{The Fit result of the $\pi^{0}\pi^{0}\eta$ mass spectrum after requiring $a_{0}(980)$ region.
    The black points with error bars represent data, and the blue solid line is the total fit.
    The red dashed line and green dashed line describes the $X(2370)$ signal and continuum background, respectively.}
\label{fig:fit_pi0pi0eta_a0}
\end{figure}

An analogous strategy to that described for the $J/\psi\to\gamma K^{0}_{S}K^{0}_{S}\pi^{0}$ channel is employed in the unbinned maximum-likelihood fit to the invariant mass spectrum of $\pi^{0}\pi^{0}\eta$ between 2.0 and $2.7~\GeV/c^{2}$, focusing on the $a_{0}(980)$ signal region, as shown in Figure~\ref{fig:fit_pi0pi0eta_a0}.
In particular, the phase space factor incorporated into this signal model corresponds to
the process $J/\psi \to \gamma 0^{-+} \to \gamma a_{0}(980)\pi^{0}$.

The systematic uncertainties for this channel are evaluated,
After evaluating all systematic uncertainties, including those sources considered in the analysis of the $J/\psi\to\gamma K^{0}_{S}K^{0}_{S}\pi^{0}$ channel,
the statistical significance of the $X(2370)$ is much greater than $5\sigma$.
The mass and width of the $X(2370)$ are determined to be
$M_{X(2370)} = 2352\pm3({\rm stat})\pm74({\rm syst})$ and $\Gamma_{X(2370)} = 134\pm 4({\rm stat})\pm62({\rm syst})$, respectively.

\begin{figure}[htbp]
    \centering
    \includegraphics[width=0.49\textwidth]{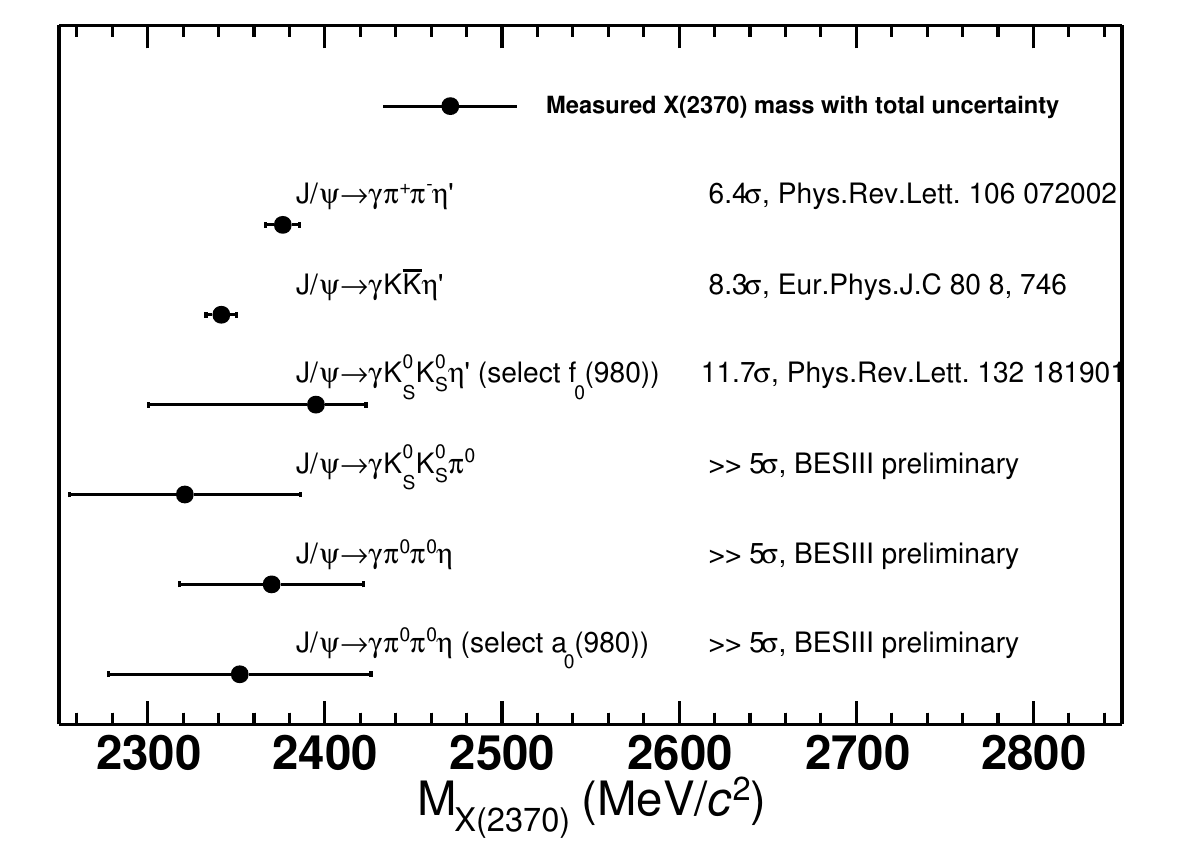}
    \includegraphics[width=0.49\textwidth]{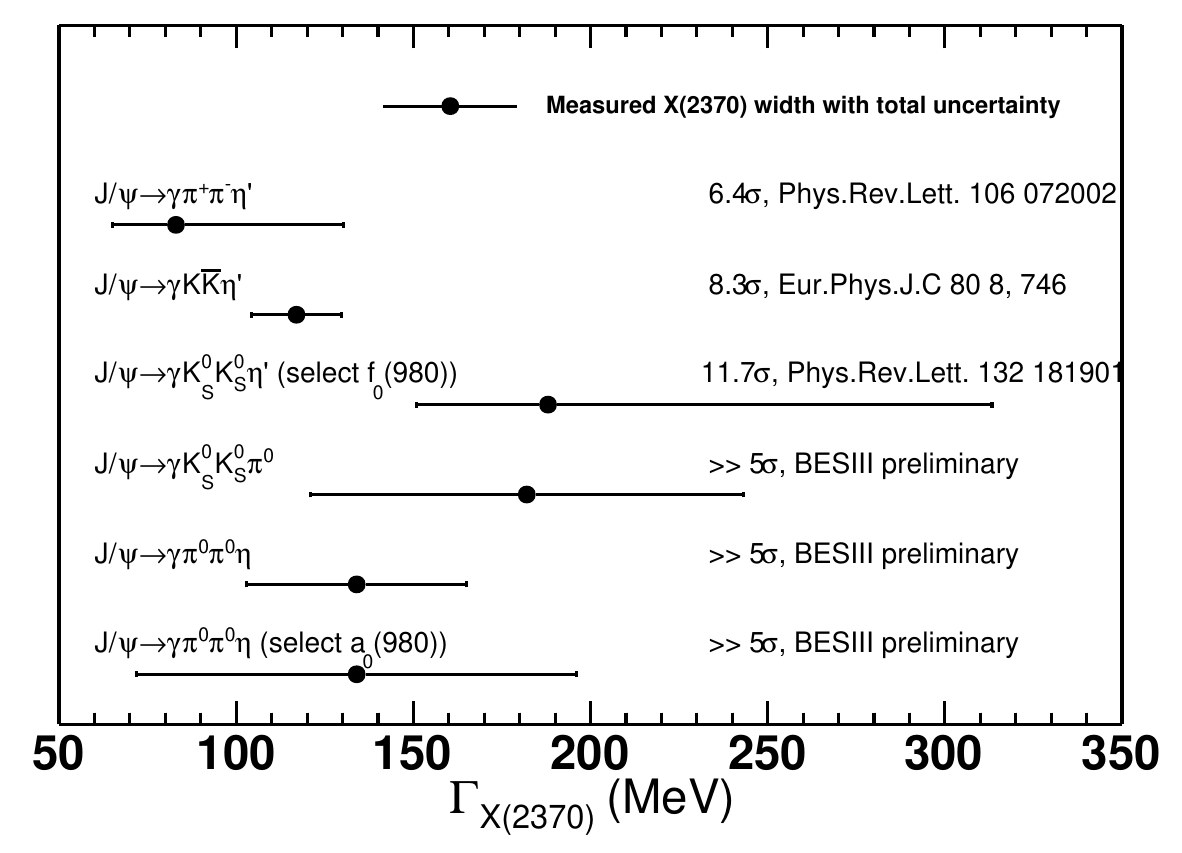}
    \caption{Summary of current observations of the $X(2370)$.}
    \label{fig:summary2370}
\end{figure}

Table~\ref{tab:NewFitResultX2370} summarizes the measurement results for the newly observed decay modes of $X(2370)$ in the $J/\psi\to\gamma K^{0}_{S}K^{0}_{S}\pi^{0}$ and $J/\psi\to\gamma \pi^{0}\pi^{0}\eta$ channels.
A comparison between new and previous observations is shown in Figure~\ref{fig:summary2370}.

\begin{table}[htbp]
\centering
 \caption{
 The measured mass, width and statistical significance of the $X(2370)$ in different decay modes.
}

\resizebox{\columnwidth}{!}{
\begin{tabular}{lccc}

\hline\hline
Decay mode     & $M_{X(2370)}$ ($\MeV/c^{2}$)            & $\Gamma_{X(2370)}$ (MeV)  & Significance \\
\hline

$X(2370) \to K_{S}^{0}K_{S}^{0}\pi^{0}$          & $2321\pm4({\rm stat})\pm65({\rm syst})$   & $182\pm 16({\rm stat})\pm59({\rm syst})$ & $\gg 5\sigma$ \\
$X(2370) \to \pi^{0}\pi^{0}\eta$          & $2370\pm2({\rm stat})\pm52({\rm syst})$   & $133\pm 8({\rm stat})\pm30({\rm syst})$ & $\gg 5\sigma$  \\
$X(2370) \to a_{0}(980) \pi^{0}$          & $2352\pm3({\rm stat})\pm74({\rm syst})$   & $134\pm 4({\rm stat})\pm62({\rm syst})$ & $\gg 5\sigma$ \\

\hline\hline
\end{tabular}
}

\label{tab:NewFitResultX2370}
\end{table}

\section{Discussion on $X(2370)$ decay and production properties}

In all 5 golden $PPP$ decay modes ($\pi\pi\eta^{\prime}$, $K\bar{K}\eta^{\prime}$, $\pi\pi\eta$, $K\bar{K}\eta$, $K\bar{K}\pi$), high similarities between the $X(2370)$ and $\eta_{c}$ decays are clearly observed, as expected for the pseudoscalar glueball decays.

In contrast, all these five decay modes with different quark flavor combinations in the final states can hardly be explained by normal $q\bar{q}$ mesons, multi-quark states or hybrid states.
For example, if the $X(2370)$ were interpreted as a normal $q\bar{q}$ meson, a large mixing between $u\bar{u}+d\bar{d}$ and $s\bar{s}$ would be needed in its $q\bar{q}$ content.
However, the LQCD calculation already showed that such a mixing for a pseudoscalar meson around $2~\GeV/c^{2}$ should be very small~\cite{Dudek:2011tt}.

\begin{figure}[htbp]
\centering

\subfloat{
  \includegraphics[width=0.4\textwidth]{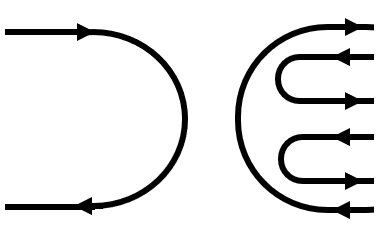}
  \label{fig:ozi_forbidden}
\put(-140, 65){(a)}
}
\subfloat{
  \includegraphics[width=0.4\textwidth]{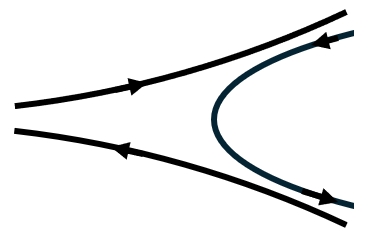}
  \label{fig:ozi_allowed}
\put(-140, 65){(b)}
}

\caption{
(a) and (b) represent the decay channels that are either suppressed by the OZI rule or not, exemplified by the process $J/\psi\to\pi^0\pi^+\pi^-$ and $\rho^0\to \pi^+\pi^-$.}
\label{fig:ozi}
\end{figure}

The narrow partial widths of these five $PPP$ decays modes of the $X(2370)$ may further disfavor the interpretations of normal $q\bar{q}$, multi-quark and hybrid states. From the BESIII observations, there is no dominant decay mode in these five modes, so they might all have similar decay branch fraction smaller than $5-10\%$ for each decay mode, then each of their partial decay widths may be even smaller than $5-10~\MeV$, which is very narrow for strong interactions. Such narrow partial decay widths can hardly be explained
if there were quark content inside $X(2370)$, including $q\bar{q}$, multi-quark and hybrid states, since its decays would be OZI allowed decays~\cite{Okubo:1963fa,Zweig:1964ruk,Zweig:1964jf,Iizuka:1966fk} and the typical OZI allowed partial decay widths are around or larger than $100~\MeV$
if there are no specific suppression factors such as phase space, and there should exit dominant decay modes for OZI allowed decays.
For example, $\rho\to\pi\pi$ process are OZI allowed decays (Figure~\ref{fig:ozi_allowed}), the partial decay width of $\rho\to\pi\pi$ is about $150~\MeV$,
and $\mathcal{B}[\rho\to\pi\pi]$ is almost 100\%.
The narrow partial decay widths and no dominant decay modes are typical signatures of OZI forbidden decays (Figure~\ref{fig:ozi_forbidden}), i.e., the decays are via gluons, just as $\eta_{c}$ decays.

Therefore the $X(2370)$ decay properties observed by BESIII experiment, i.e., observation of the $X(2370)$ in all 5 golden $PPP$ decay modes with narrow partial decay widths and without dominant decay modes, strongly favor the glueball interpretation of the $X(2370)$.

In terms of the $X(2370)$ production property, the high production rate in $J/\psi$ radiative decays is just the expectation for a glueball.
From $\mathcal{B}[J/\psi\to \gamma X] \times \mathcal{B}[X\to f_0(980)\eta^{\prime}] \times \mathcal{B}[f_0(980)\to K^{0}_{S}K^{0}_{S}] = \left( 1.31 \pm 0.22 ({\rm stat})^{+2.85}_{-0.84}({\rm syst}) \right) \times 10^{-5}$, assuming $\mathcal{B}[X(2370)\to f_0(980)\eta^{\prime}]$ is about 5\%, one can obtain $B[J/\psi\to \gamma X]$ is about $10.7^{+22.8}_{-7}\times 10^{-4}$, which is consistent with LQCD calculation~\cite{2370_prediction_glueball_2019} for the $0^{-+}$ glueball $(2.31\pm0.80)\times10^{-4}$.

In the mass region above $2.3~\GeV/c^{2}$ as the $0^{-+}$ glueball mass region from the LQCD prediction, the $X(2370)$ is the unique $0^{-+}$ particle produced in all 5 golden decay modes in $J/\psi$ radiative decays. Then no other $0^{-+}$ particle can be called as ``richly produced'' in $J/\psi$ radiative decays if they have not been observed in 10 billion such a huge $J/\psi$ data sample. So the requirement on high production rate in $J/\psi$ radiative decays would basically exclude all other potential $0^{-+}$ glueball candidates in mass region above $2.3~\GeV/c^{2}$.

\section{Conclusion and outlook}

In summary, BESIII experiment discovers a glueball-like particle $X(2370)$\textemdash the mass, spin-parity quantum numbers, production and decay properties are fully consistent with the features of the lightest pseudoscalar glueball. So far, only the glueball interpretation can naturally explain all of the decay and production properties without any difficulties or contradictions.

More decay modes of $X(2370)$ will be searched for and studied with 10 billion $J/\psi$ data sample. Among them, the $K^*K$ decay mode is of special importance since it should be suppressed in a $0^{-+}$ flavor singlet decays \cite{Lipkin:1981uc,Lipkin:1981ak,Klempt:2007cp} (such as a $0^{-+}$ glueball and $\eta_{c}$ decays). So searching for the $K^*K$ decays of $X(2370)$ in $J/\psi\to \gamma K^*K\to \gamma K\bar{K}\pi$ will provide a crucial test whether $X(2370)$ is a flavor singlet state as an expectation for a glueball.

\section*{Acknowledgments}
The authors would like to thanks Prof. Beijiang Liu for his helpful discussions.
The BESIII Collaboration thanks the staff of BEPCII and the IHEP computing center for their strong support.

\bibliographystyle{ws-ijmpa}
\bibliography{sample}

\end{document}